\documentclass[vecphys]{svmult}

\RequirePackage{epsfig}

\usepackage{makeidx}         
\usepackage{graphicx}        
\usepackage{multicol}        
\usepackage[bottom]{footmisc}
\usepackage{amsfonts}                                                 

\makeindex             

\begin{document}

\title*{Magnetohydrodynamics of Protostellar Disks}

\author{Steven A. Balbus}
\institute{{Laboratoire de Radio Astronomie, \'Ecole Normale Sup\'erieure,
24 rue Lhomond, 75231 Paris EDEX 05, France}
\texttt{balbus@ens.fr}        }
%
%
\maketitle

The magnetohydrodynamical behavior (MHD) of accretion disks is
reviewed.  A detailed presentation of the fundamental MHD equations
appropriate for protostellar disks is given.  The combination of a weak
(subthermal) magnetic field and Keplerian rotation is unstable to the
magnetorotational instability (MRI), if the degree of ionization in the
disk is sufficiently high.  The MRI produces enhanced angular momentum
and leads to a breakdown of laminar flow into turbulence.  If the
turbulent energy is dissipated locally, standard ``$\alpha$'' modeling
should give a reasonable estimate for the disk structure.  Because away
from the central star the ionization fraction of protostellar disks is
small, they are generally not in the regime of near perfect conductivity.
Nonideal MHD effects are important.  Of these, Ohmic dissipation and Hall
electromotive forces are the most important.  The presence of dust is
also critical, as small interstellar scale grains absorb free charges
that are needed for good magnetic coupling.  On scales of AU's there
may be a region near the disk midplane that is magnetically decoupled,
a so-called {\em dead zone.} But the growth and settling of the grains as
time evolves reduces their efficiency to absorb charge.  With ionization
provided by coronal X-rays from the central star (and possibly also cosmic
rays), protostellar disks may be sufficiently magnetized throughout most
of their lives to be MRI active, especially away from the disk midplane.

\newcommand\bb[1] { \mbox{\boldmath{$#1$}} }
\newcommand\del{\bb{\nabla}}
\newcommand\bcdot{\bb{\cdot}}
\newcommand\btimes{\bb{\times}}
\newcommand\vv{\bb{v}}
\newcommand\ve{\bb{v_e}}
\newcommand\ven{\bb{v_e} -\vv}
\newcommand\vi{\bb{v_I}}
\newcommand\vin{\bb{v_I}-\vv}
\newcommand\pen{\bb{p_{en}}}
\newcommand\pin{\bb{p_{In}}}
\newcommand\bg{\bb{g}}
\newcommand\rr{\bb{r}}
\newcommand\eps{\epsilon}
\newcommand\epsijk{\epsilon^{ijk}}
\newcommand\B{\bb{B}}
\newcommand\J{\bb{J}}
\newcommand\EE{\bb{E}}
\newcommand\BV{Brunt-V\"ais\"al\"a\ }
\newcommand\iw{ i \omega }
\newcommand\kva{ \bb{k\cdot v_A}  }
\newcommand\beq{\begin{equation}}
\newcommand\eeq{\end{equation}}

\def\dd{\partial}

\section{Introduction}
\label{sec:1}

An understanding of the magnetohydrodynamics (MHD) of protostellar
disks is crucial for the theoretical development of these objects.
There is no getting around the fact that the subject is decidedly
inelegant.  In principle, we are only interested in solving the equation
$\bb{F}=m\bb{a}$, but the range of topics that bears on this problem
is truly daunting.  Molecular chemistry, dust grain physico-chemistry,
photo-ionization physics, aerosol theory, and non-ideal MHD are all
key players in this game.  If the results of the theorists' efforts had
been little or no progress, there would have been no shortage of excuses.
(``More work is needed.'')   But in fact there has been substantial progress
in understanding important issues, more perhaps than some practitioners
may even realize.   A sort of consensus is beginning to take shape
of the gross properties of protostellar disks that
are dictated by the demands of MHD physics.  In this review, I will try
to put forward as strong a case as I can for what I somewhat boldly
regard as the canonical protostellar disk model, while at the same time
try not to gloss over what are genuine uncertainties or difficulties.

Protostellar disks are a class of accretion disks, and one area where
there certainly {\em has} been a great deal of progress in recent years
is in the development of accretion disk theory.  The realization that
a combination of differential rotation and a weak magnetic field is
profoundly unstable and produces turbulence has given the subject a 
foundation on which to build (Balbus \& Hawley 1991, 1998).   Indeed,
the primary reason for studying MHD processes in protostellar disks
in detail is that once these systems are no longer self-gravitating,
magnetic fields profoundly influence their dynamical behavior.  It is
not possible to understand angular momentum transport in low mass
(``T-tauri'') disks without focusing on magnetic fields.

Magnetic fields provide a conduit for free energy to flow from the
differential rotation to the disk itself, producing fluctuations,
turbulent heating, and quite possibly the directly observed
outflows.  Most importantly, we shall see that MHD turbulence produces
well-correlated fluctuations of the radial and azimuthal components of
the magnetic field and fluctuating gas velocity.  It is precisely these
correlations that result in a substantial outward transport of angular
momentum, allowing the disk gas to spiral inward and accrete onto the
central star.  

It may also be possible in principle for winds to remove angular
momentum from the disk without completely disrupting it.  (See Koenigl,
this volume.)  The difficulty that needs to be overcome for this mechanism
is that the outflow must occur over the whole face of the disk, not just
from the innermost regions from where jets are launched: material near
the inner disk edge has already lost almost all of its angular momentum.
Instead, the outer disk material, torqued by the magnetic field, must
continuously slip relative to the field, or else the field tends to
become very centrally concentrated (Lubow, Papaloizou, \& Pringle 1994).
How this might work without generating internal disk turbulence that
itself transports angular momentum remains to be fully understood.


This understanding that magnetic fields play a crucial role as the
source of disk turbulence developed more than thirty years after the
basic instability (now known as the magnetorotational instability
or MRI) first appeared in the literature (Velikhov 1959).  But even
if the importance of the instability had been immediately grasped,
without the computational power that became available only post 1990,
it is doubtful that the impact would have been the same.   There is no
substitute for being able to visualize on your desktop the development
of a linear instability into full-fledged MHD turbulence.

The MRI depends upon the presence of electrical currents to do its job,
and this means that the disk gas must be at least partially ionized.
``Partially ionized'' in practice could mean even a minute electron
fraction.  Tiny traces of electrons can magnetize the disk, an effect we
will quantify in \S 8.  It is because even a wisp of a magnetic field
and the merest trace of electrons take on such dynamical significance
that protostellar disk dynamics depends so heavily on protostellar
disk chemistry.

Young protostellar disks need sodium, potassium and other trace
``vitamins'' to make them vigorous and active, because these alkalis
are important in regulating the delicate ionization balance in the gas.
It is ultimately the ionization level that determines the resistivity that
determines whether the magnetic field is well-coupled to the gas or not.
This is not an issue in accretion disks around compact objects, for
which even a small fraction of the turbulent energy dissipated would be
enough to ensure that the gas is thermally ionized.  Protostellar disks,
on the other hand, are big (so the free energy of differential rotation
is small), dusty (so grains absorb free electrons), and cold (so thermal
ionization is unimportant except near the star).  The overarching
question of protostellar MHD research is to understand under what
conditions the abundance of free electrons level drops below the level
needed to ensure good magnetic coupling.  The magnetic coupling leads
in turn to instability, turbulence, and enhanced transport, the essence
of disk dynamics.  Much of this review will be devoted directly to the
question of disk ionization balance.  For readers wishing additional
astronomical background material, the text of Lee Hartmann (1998) is
an excellent choice.  Balbus \& Hawley (2000) and Stone et al. (2000)
are earlier reviews of MHD transport processes in protostellar disks.
More general disk reviews include Pringle (1981; a classic review
but pre-MRI), Papaloizou \&
Lin (1995), Lin \& Papaloizou (1996), Balbus \& Hawley (1998), and Balbus
(2003).   In addition, the {\em Protostars and Planets} series
published by the University of Arizona Press series contains a useful
historical record of the development of the subject of protostellar
disks; the latest volume is {\em Protostars and Planets V} (Reipurth,
Jewitt, \& Keil 2007).

\section {On the Need for MHD}

The modern view of protostellar disks is heavily influenced by accretion
disk formalism.  Lin \& Papaloizou (1980) is the first pioneering study to
bring to bear accretion disk formalism to the study of protostellar disks,
though it is now thought that MHD turbulence, rather than convective
turbulence, is the key to protostellar disk dynamics.  To understand the
central role of a magnetic field in our understanding of protostellar
disks, it is of interest to review where we stand with respect to our
knowledge of the onset of turbulence in rotating flows.

The principle problem for classical disk theory (Shakura \& Sunyaev 1973) was
to discover the origin of the very large stresses that were needed to
transport angular momentum, a process then (and still often now) referred
to as ``anomalous viscosity.''  The approach of Shakura \& Sunyaev was to
make a virtue of necessity, arguing that the small microscopic viscosity
implied a very large Reynolds number (Re)\footnote{The Reynolds number
Re is the ratio of a characteristic flow velocity $V$ times a charateristic
length $L$ divided by the kinematic viscosity, $\nu$, ${\rm Re}=VL/\nu$.},
and a large Reynolds number
meant that shear turbulence was present.  This conclusion was sustained
by the belief that the laboratory experiments showing a nonlinear
breakdown of Cartesian shear layers at values of Re in excess of $\sim
10^3$ would naturally carry over to Keplerian disks, where its value was
considerably higher.  The fact that the inertial force associated with
the Coriolis effect (not present in planar flow) was well in excess of
the destabilizing shear does not seem to have been viewed as troublesome
to proponents of this view.

The classical laboratory method for investigating the stability
of rotating flows is a Couette apparatus.  In such a device, the
space between two coaxial cylinders is filled with a liquid, almost
always ordinary water.  The two cylinders rotate at different angular
velocities, let us say $\Omega_{in}$ and $\Omega_{out}$.  The gap between
the cylinders is of order a centimeter, and a stable rotation profile
will be attained in a matter of minutes by viscous diffusion, even if
Re is very large.  (An {\em unstable} rotation profile will of course
never be found as such; the flow will remain permanently disturbed.)
By choosing $\Omega_{in}$ and $\Omega_{out}$ appropriately, a section of
a Keplerian disk can be accurately mimicked, and the effects of Coriolis
forces can be studied.

According to the classic text of Landau \& Lifschitz (1959), a Couette
rotation profile that is found in the laboratory to be stable ``does not
actually mean, however, that the flow actually remains steady no matter
how large [Re] becomes.''   The monograph by Zel'dovich, Ruzmaikin, \&
Sokoloff (1983) is even more explicit on the question, unambiguously
stating that laboratory experiments had already shown that Keplerian
rotation profiles were nonlinearly unstable at sufficiently large Re.

These are stunning claims.  At the time these statements were written,
there was certainly no credible laboratory evidence that Keplerian
rotation profiles were nonlinearly unstable.  The first explicit claim
that Keplerian flow was unstable based on laboratory evidence seems to
be the unpublished result of Richard (2001).  However, a later experiment
by Ji et al. (2006) suggests that the earlier finding of instability was
probably due to spurious effects arising from the interaction between
the flow and the endcaps of the experiments (Ekman layers\footnote{
Ekman layers are narrow fluid layers between a solid boundary and a
rotating flow.  When the bulk of the flow is not at rest in the frame of
the boundary, the fluid rotation changes rapidly in the Ekman layer and
can produce a secondary circulation pattern invading the entire flow.}).
When care is taken to minimize this interaction, Keplerian flow is found
to be linearly and nonlinearly stable at values of Re up to $2\times
10^6$.  Serious students of disk theory would do well to develop a deep
skepticism of ``large Re means disks are turbulent'' arguments, which
sadly are still promulgated in contemporary textbooks.

Why is the Coriolis force is so harmful to the maintenance of turbulence?
Moffatt, Kida, \& Ohkitani (1994) have referred to stretched vortices
as the ``sinews of turbulence.''  Turbulence is maintained in shear
flows by vortices that are ensnared along the axis of strain, coupling
the free energy of the shear directly to the internal vortex motion.
Two neighboring fluid elements in a vortex are rapidly pulled apart as
its circulation rises.  This is possible only if Coriolis forces are
absent.  Despite the presence of shear, Coriolis forces induce epicyclic,
oscillatory motions that do not allow anything resembling continuous
vortex stretching.  Without this, turbulence cannot be maintained.  This
argument breaks down of the strain rate much exceeds the formal epicyclic
oscillation period, even the latter is quite well defined.  Such flows,
in fact, are found to be nonlinearly unstable in numerical simulations.
But for Keplerian flow, the epicyclic frequency is just the local $\Omega$
of the disk, and it exerts a strongly stabilizing influence.  

Figure (1) illustrates this point.  Locally, Cartesian shear flow and
Keplerian differential rotation appear to be similar.  But perturbed
fluid elements behave very differently in the two systems.  Viewed from
a frame moving with the undisturbed flow, a displaced fluid element in
shear flow approximately follows an unbounded parabolic trajectory A,
while a displaced fluid element in a Keplerian disk approximately follows
a bound epicycle, B.   A small perturbing vortex would be stretched
continuously in shear flow, tapping into the free energy source needed
to maintain turbulence.  By contrast, the embedded vortex would merely
oscillate in a disk.  It is this difference, due entirely to the presence
of the rotational Coriolis force, that is ultimately responsible for the
hydrodynamical stability of Keplerian disks, whereas shear flow can be
nonlinearly unstable.

\begin{figure}
\epsfig{file=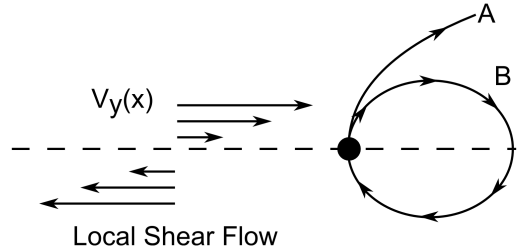, width=10cm}
\caption {The stabilization mechanism in a rotating disk versus Cartesian
shear flow.  In shear flow, displacements are
unbounded, as indicated by path A.  In a disk, epicyclic
motions keeps the displacement tightly bound. See text for
further details.} \end{figure}

The title of the Ji et al. (2006) paper is ``Hydrodynamic turbulence
cannot transport angular momentum effectively in astrophysical disks,''
and that seems as good a one-sentence summary on the topic as any;
their experiment is probably the final word on nonlinear Keplerian
hydrodynamical shear instability.   It is gratifying, therefore, that
all of the difficulties encountered by Coriolis stabilization vanish
when magnetic fields are taken into account.  This is due to the fact
that the magnetic field introduces new modes of response (shear waves)
that are much less prone to Coriolis stabilization.  One of these
waves, the so-called ``slow mode,'' turns into a local instability when
differential rotation is present with a weak magnetic field.  This is
the magnetorotational instability.  To understand this porcess in more
detail, it is best to begin with the fundamentals of MHD.

\section {Fundamentals}

In this section, a detailed derivation of the fundamental MHD equations
is presented.  The discussion will be more technical here than in most
of the rest of the paper, but it is very important to see how the basic
governing equations of the subject arise, and much of this material
is not so easy to find outside of specialized treatments.  I hope the
reader will have the patience to read carefully through this section.

A protostellar disk is a gas of neutral particles (predominantly
$H_2$ molecules), electrons, ions (the most important of which
will generally be K$^+$ and Na$^+$), and dust grains.  We defer to
\S 8 a discussion of the complications due to the presence of dust
grains and consider here the gas dynamical equations for mixture of
neutrals, ions and atoms.

Each species (denoted by subscript $s$) is separately conserved,
and obeys the mass conservation equation
\beq\label{mass1}
{\dd\rho_s\over \dd t} + \del\bcdot (\rho_s\vv_s) = 0
\eeq
where $\rho_s$ is the mass density for species $s$ and $\vv_s$
is the velocity.  The flow quantities for the dominant neutral
species will henceforth be presented without subscripts.

The dynamical equation for the neutral particles is
\beq
\rho {\dd\vv\over\dd t} + \rho (\vv\bcdot\del)
\vv = -\del P
-\rho\del\Phi  -\bb{p_{nI}} - \bb{p_{ne}}
\eeq
where $P$ is the pressure of the neutrals, $\Phi$ the gravitational
potential and $\bb{p_{nI}}$ ($\bb{p_{ne}}$) is the momentum exchange
rate between the neutrals and the ions (electrons).

Let us examine these last two important terms in more detail.
$\bb{p_{nI}}$ takes the form
\beq
\bb{p_{nI}} = n \mu_{nI} (\vv -\bb{v_I}) \nu_{nI}
\eeq
where $n$ is the number density of neutrals,
and $\mu_{nI}$ is the reduced mass of an ion and neutral particle,
\beq
\mu_{nI} \equiv {m_I m_n\over m_I +m_n},
\eeq
$m_I$ and $m_n$ being the ion and neutral mass respectively.
$\nu_{nI}$
is the collision frequency of a neutral with a population of ions,
\beq\label{nuni}
\nu_{nI} = n_I \langle \sigma_{nI} w_{nI}\rangle.
\eeq
In equation (\ref{nuni}), $n_I$ is the number density of ions,
$\sigma_{nI}$ is the effective cross section for neutral-ion collisions,
and $w_{nI}$ is the relative velocity between a neutral particle and
an ion.  The angle brackets represent an average over a Maxwellian
distribution function for the relative velocity.  (The mass appearing
in this Maxwellian will of course be the reduced mass
$\mu_{nI}$.)  For neutral-ion scattering,
we may take the cross section $\sigma_{nI}$ to be approximately geometrical,
which means that the quantity in angle brackets will be proportional to
$\mu_{nI}^{-1/2}$.  The order of the subscripts
has no particular significance for the cross section $\sigma_{nI}$
reduced mass $\mu_{nI}$ or 
relative velocity $w_{nI}$.   But $\nu_{In}$ differs from
$\nu_{nI}$: the former is proportional
to the neutral density $n$, not the ion density $n_I$.  

Putting these definitions together gives
\beq
\bb{p_{nI}} = n n_I \mu_{nI}
\langle \sigma_{nI} w_{nI}\rangle(\vv -\bb{v_I})
\eeq
In accordance with Newton's third law, 
this is symmetric with respect to the interchange $n\leftrightarrow I$,
except for a change in sign, $\bb{p_{nI}}=-\bb{p_{In}}$.
All of these considerations hold, of course,
for electron-neutral scattering
as well.  Explicitly, we have
\beq
\bb{p_{ne}} = n n_e \mu_{ne}
\langle \sigma_{ne} w_{ne}\rangle(\vv -\bb{v_e})
\simeq n n_e m_e
\langle \sigma_{ne} w_{ne}\rangle(\vv -\bb{v_e}).
\eeq
The gas will be locally neutral, so that $n_e=Zn_i$ where $Z$ is
the number of ionizations per ion particle.   In a weakly ionized gas,
$Z=1$.  The reduced mass $\mu_{ne}$ may safely be set
equal to the electron mass $m_e$.
We will use the following expressions for the collision rates
(Draine, Roberge, \& Dalgarno 1983)\footnote{A recent calculation
of H-H$^+$ scattering by Glassgold, Krsti\`c, \& Schultz (2005)
may imply a slightly higher value for the neutral-ion collision rate.}:
\beq
\langle \sigma_{nI} w_{nI}\rangle = 1.9\times 10^{-9} \mbox{ cm$^3$ s$^{-1}$}
\eeq
\beq
\langle \sigma_{ne} w_{ne}\rangle = 10^{-15}\ (128kT/9\pi
m_e)^{1/2} = 8.3\times 10^{-10} T^{1/2} \mbox { cm$^3$ s$^{-1}$}
\eeq
The electron-neutral collision rate is essentially
the ion geometric cross section times an electron thermal
velocity.  (The peculiar factor of $(128/9\pi)^{1/2}$
comes from the way the electron velocity is projected,
when averaged over the electron Maxwellian distribution function,
along the direction of the mean ion flow.)
But the ion-neutral collision rate is temperature independent,
much more beholden to long range induced dipole interactions,
and significantly enhanced relative to a geometrical cross section assumption.
Even if the ion-neutral rate were determined only by a geometrical
cross section, 
$|\bb{p_{nI}}|$ would excee $|\bb{p_{ne}}|$ by a factor of order
$(m_e/\mu_{nI})^{1/2}$.  In fact, the dipole
enhancement of the ion-neutral cross section makes this
factor yet larger.

In the astrophysical literature,
it is common to write the ion-neutral momentum coupling in the
form
\beq\label{pinn}
\pin = \rho\rho_I \gamma (\vin),
\eeq
where $\gamma$ is the so-called {\em drag coefficient},
\beq
\gamma \equiv { \langle \sigma_{nI} w_{nI}\rangle\over m_I+m_n}
\eeq
and we will use this notation from here on.
Numerically, $\gamma=3\times 10^{13}$ cm$^3$ s$^{-1}$ g$^{-1}$
for astrophysical mixtures (Draine, Roberge, \& Dalgarno 1983).

The dynamical equations for the ions and electrons are
\beq
\rho_I{\dd\vi\over\dd t} + \rho_I \vi\bcdot\del\vi = -\del P_I -\rho_I\del\Phi
+Zen_I\left(
\bb{E} + {\bb{\vi}\over c}\btimes\bb{B}
\right) - \bb{p_{In}}
\eeq
and
\beq
\rho_e{\dd\ve\over\dd t} + \rho_e \ve\bcdot\del\ve = -\del P_e -\rho_e\del\Phi
-en_e\left(
\bb{E} + {\bb{\ve}\over c}\btimes\bb{B}
\right) - \bb{p_{en}},
\eeq
respectively.  Throughout this paper, $e$ will denote
the {\em positive} charge of a proton, $4.803\times 10^{-10}$ esu.
For a weakly ionized gas, the Lorentz force and collisional
terms dominate in each of the latter two equations.  Comparison of
the magnetic and inertial forces, for example,
shows that the latter are smaller than
the former by the ratio of the proton or electron
gyroperiod to a macroscopic flow crossing time.
Thus, to an excellent degree of approximation,
\beq
Zen_I\left(
\bb{E} + {\bb{\vi}\over c}\btimes\bb{B}\right) - \bb{p_{In}}
=0,
\eeq
and
\beq
-e n_e\left(
\bb{E} + {\bb{\ve}\over c}\btimes\bb{B}\right)- \bb{p_{en}}
=0.
\eeq
The sum of these two equations gives
\beq\label{ambi}
{ \bb{J}\over c}\btimes \bb{B} = \bb{p_{In}} + \bb{p_{en}}
\eeq
where charge neutrality $n_e=Zn_I$ has been used,
and we have introduced the current density
\beq\label{J}
\bb{J} \equiv en_e(\vi-\ve).
\eeq
The equation for the neutrals becomes
\beq\label{neuts}
\rho {\dd\vv\over\dd t} + \rho \vv\bcdot\del \vv = -\del P
-\rho\del\Phi  + { \bb{J}\over c}\btimes \bb{B}
\eeq
Due to collisional coupling, the neutrals are subject to the magnetic
Lorentz force just as though they were a gas of charged particles.  It is
not the magnetic force {\em per se} that changes in a neutral gas.
As well shall presently see, it is the inductive properties of the gas.

Let us return to the force balance equations for the electrons:
\beq
-e n_e\left(
\bb{E} + {\bb{\ve}\over c}\btimes\bb{B}\right)- \bb{p_{en}}=0.
 \eeq
After division by $-en_e$, this may be expanded to
 \beq\label{vees}
 \bb{E}+
 {1\over c} \left[
 \vv +(\ve-\vi)+(\vi-\vv)\right] \btimes\bb{B} + {m_e \nu_{en}\over e}
 \left[ (\ve-\vi) +(\vi -\vv)\right]=0,
 \eeq
where we have introduced the collision frequency of an electron in
a population of neutrals:
\beq
\nu_{en} =n\langle \sigma_{ne} w_{ne} \rangle.
\eeq
We have written the electron velocity $\ve$ in
terms of the dominant neutral
velocity $\vv$ and the key physical velocity differences of
our problem.  It has already been
noted that in equation (\ref{ambi}),
$\bb{p_{en}}$ is small compared with $\bb{p_{In}}$,
provided that the velocity difference $|\ve-\vv|$ is not
much larger than $|\vi-\vv|$.  In fact, when used in (\ref{vees}),
the $\bb{p_{en}}$ term
in equation (\ref{ambi}) may quite
generally be dropped:
 \beq
 { \bb{J}\over c}\btimes \bb{B} \simeq \bb{p_{In}}.
 \eeq
It may also be shown that
the final term in equation (\ref{vees})
 $$
 {m_e \nu_{en}\over e}(\vi -\vv),
 $$
which would then be proportional
to $\bb{J\times B}$, becomes small compared with the
third term
 $$
 {1\over c} (\ve-\vi) \btimes\bb{B},
 $$
also proportional to $\bb{J}\btimes\bb{B}$.
Both of these rather technical claims are justified in detail in Appendix A.
(In both cases, the incurred error is of order $(m_e/\mu_{In})^{1/2}$.)
These simplifications allow us to write the electron force balance equation
as
 \beq\label{ohm1}
 \bb{E} +{\vv\over c}\btimes \bb{B} -{ \bb{J\times B}\over en_e c}
 -{ \bb{J}\over \sigma_{cond}} + { \bb{ (J\times B)\times B}\over
 c^2\gamma\rho \rho_I} = 0,
 \eeq
where the electrical conductivity has been defined as
 \beq\label{cond1}
 \sigma_{cond} \equiv {e^2n_e\over m_e \nu_{en}}
 \eeq
The associated resistivity $\eta$ is (e.g. Jackson 1975)
 \beq\label{res1}
 \eta = {c^2\over 4\pi\sigma_{cond}},
 \eeq
which has units of cm$^2$ s$^{-1}$. Numerically (e.g. Blaes \& Balbus
1994, Balbus \& Terquem 2001):
\beq
\eta = 234 \left(n \over n_e\right) T^{1/2}\ \mbox{cm$^2$ s$^{-1}$}
\eeq
Equation (\ref{ohm1}) is a general form of Ohm's law for a moving,
multiple fluid system.

Next, we make use of two of Maxwell's equations.  The first
is Faraday's induction law:
 \beq\label{faraday}
 {\del\btimes\bb{E}}= -{1\over c}{\dd\bb{B}\over \dd t}.
 \eeq
We substitute $\bb{E}$ from equation (\ref{ohm1})
to obtain an equation for the self-induction of the magnetized
fluid,
 \beq\label{ind1}
 {\dd\bb{B}\over \dd t}=
 \del\btimes\left[
 {\vv}\btimes \bb{B} -{ \bb{J\times B}\over en_e }
 +
 { \bb{ (J\times B)\times B}\over
 c\gamma\rho \rho_I}
 - { c\bb{J}\over \sigma_{cond}}\right]
 \eeq

It remains to relate the current density $\J$ to the magnetic
field $\B$.  This is accomplished by the second Maxwell equation,
 \beq\label{displace}
 {4\pi\over c}\J = \del\btimes\B +{1\over c}{\dd\EE\over \dd t}
 \eeq
The final term in the above is the displacement current,
and it may be ignored.  Indeed,
since we have not, and will not, use the ``Gauss's Law''
equation
 \beq
 \del\bcdot\EE = 4\pi e (Zn_I-n_e),
 \eeq
we {\em must not}
include the displacement current.  In Appendix B, we show that
departures from charge neutrality in $\del\bcdot\EE$ and the displacement
current are both small terms that contribute at the same order: $v^2/c^2$.
These must both be self-consistently neglected in nonrelativisitc MHD.
(The final Maxwell equation $\del\bcdot\B=0$ adds nothing new.  It is
automatically satisfied by equation (\ref{faraday}), as long as the initial
magnetic field satifies this divergence free condition.)
These considerations imply
 \beq\label{J1}
 \J = {c\over 4\pi}\del\btimes\B
 \eeq
for use in equation (\ref{ind1}).

To summarize, the fundamental equations of a weakly
ionized fluid are mass conservation of the
dominant neutrals (eq.[\ref{mass1}])
 \beq
 {\dd\rho\over \dd t} + \del\bcdot(\rho\vv)=0,
 \eeq
the equation of motion (eq. [\ref{neuts}] with [\ref{J1}])
 \beq
 \rho {\dd\vv\over\dd t} + \rho \vv\bcdot\del \vv = -\del P
 -\rho\del\Phi  +  {1\over4\pi}(\del\btimes\B)\btimes \bb{B},
 \eeq
and the induction equation (eq. [\ref{ind1}] with [\ref{res1}]
and [\ref{J1}])
 \beq\label{full}
 {\dd\bb{B}\over \dd t}=
 \del\btimes\left[
 {\vv}\btimes \bb{B} -{ c(\bb{\nabla \times B})\times \B\over 4\pi en_e }
 +
 { \bb{ [  (\nabla\times B)\times B]\times B}\over
 4\pi \gamma\rho \rho_I}
 -  \eta \bb{\nabla\times B}\right]
 \eeq

It is only natural that the reader should be a little taken aback by the
sight of equation (\ref{full}).  Be assured that it is rarely, if ever,
needed in full generality: almost always one or more terms on the right
side of the equation may be discarded.   When only the induction term
$\vv\btimes\bb{B}$ is important, we refer to this regime as {\em ideal}
MHD.  The three remaining terms on the right are the {nonideal} MHD terms.

To get a better feel for the relative importance of the 
nonideal MHD terms in equation (\ref{ind1}), we denote
the terms on the right side of the equation, moving left to right,
as $I$ (induction), $H$ (Hall), $A$ (ambipolar diffusion), and $O$
(Ohmic resistivity).   Then, the scaling of each of these terms
relative to the Hall term $H$ is
 \beq
 {I\over H} \sim {v\over v_I-v_e},\quad {A\over H} \sim Z{\omega_{cI}\over
 \gamma\rho}, \quad {O\over H} \sim {\nu_{en}\over \omega_{ce}}
 \eeq
where the ion cyclotron frequency $\omega_{cI}=eB/m_I c$, and similarly
for the electron frequency $\omega_{ce}$, with $m_e$ replacing $m_I$.

We will always be in a regime in which the presence of the induction
term is not in question, i.e. the relative ion-electron drift velocity
$v_I-v_e$ is always comparable to or in fact much less than $v$.
More interesting is the relative importance of the nonideal terms.
The explicit dependence of $A/H$ and $O/H$ in terms of
the fluid properties of a cosmic gas is given in Balbus \& Terquem (2001):
 \beq\label{AH}
 {A\over H} = Z \left(9\times 10^{12}\ {\rm cm}^{-3}\over n\right)^{1/2}
 \left(T\over 10^3\, {\rm K}\right)^{1/2}
 \left(v_A\over c_S\right)
 \eeq
and
 \beq\label{OH}
 {O\over H} = \left(n\over 8\times 10^{17}\ {\rm cm}^{-3}\right)^{1/2}
 \left(c_S\over v_A\right)
 \eeq
Here $n$ is the total number density of all particles, $T$ is the
kinetic temperature, $v_A$ is
the so-called {\em Alfv\'en velocity},
 \beq
 \bb{v_A} = {\bb{B}\over \sqrt{4\pi\rho}}
 \eeq
and $c_S$ is the isothermal speed of sound,
\beq
 c_S^2 = 0.429 {kT\over m_p}
\eeq
where $k$ is the Boltzmann constant and $m_p$ the mass of the proton.
The coefficient 0.429 corresponds to a mean mass per particle of
2.33$m_p$, appropriate to a molecular cosmic abundance gas.

As reassurance that the fully general nonideal MHD induction
equation is not needed for our purposes, note that
equations (\ref{AH}) and (\ref{OH}) imply that for all three nonideal
MHD terms to be comparable, $T\sim 10^8$ K!  Obviously this is not an
issue for protostellar disks.  In figure (2), we plot the domains
of relative dominance of the nonideal MHD terms in the $nT$ plane.
Protostellar disks are dominated by the resistive
and Hall nonideal MHD terms, except
in the innermost regions, where ohmic dissipation is largest,
and the outermost regions, where ambipolar diffusion becomes important
(Wardle 1999).

\begin{figure}
\epsfig{file=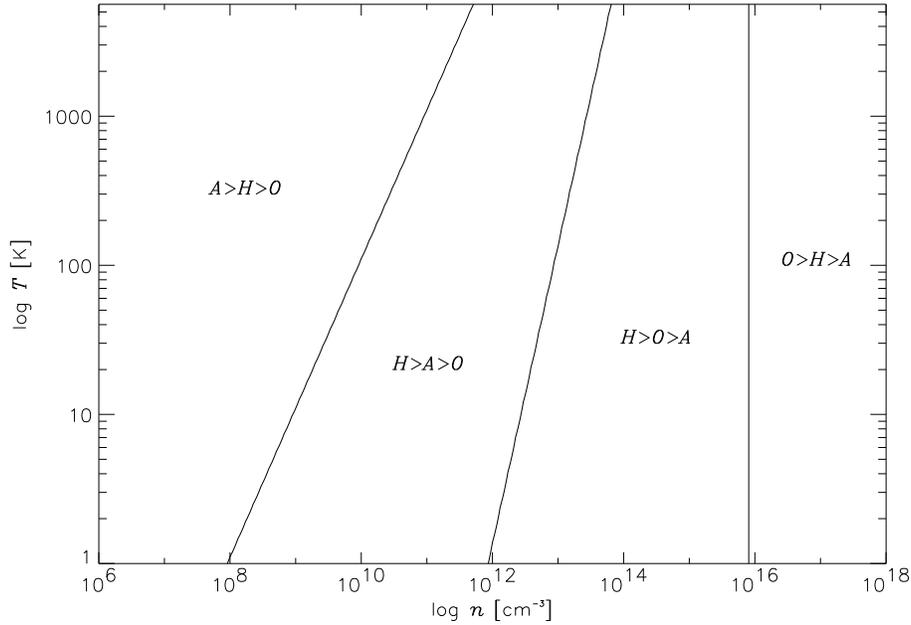, width=12cm}
\caption {Parameter space for nonideal MHD.  The curves correspond to 
the case $v_A/c_S=0.1$.
(From Kunz \& Balbus 2004.)}
\end{figure}

Our emphasis of the relative ordering of the nonideal terms in the
induction equation should not obscure the fact that 
ideal MHD is often an excellent approximation, even when the 
ionization fraction is $ \ll 1$.  For example, the ratio of the
ideal inductive
term to the ohmic loss term is given by the Lundquist number
\beq
\mathbb {L} = {v_A L\over \eta}
\eeq
where $L$ is a characteristic gradient length scale.   
To orient ourselves, let us set $L=0.1R$,
where $R$ is the radial location in the disk.  Then $\mathbb{L}$
is given by
$$
\mathbb{L} \simeq 2.5 (n_e/n) (v_A/c_S) R_{cm},
$$
$R_{cm}$ being the radius in centimeters.  
In other words, the critical ionization fraction at which 
$\mathbb{L}=1$ is about 
$$(n_e/n)_{crit}=0.4(c_S/v_A)R_{cm}^{-1}\sim 10^{-13}(c_S/10v_A)$$
at $R=1$ AU.
The actual nebular ionization
fraction at this location may be above or below this
during the course of the solar systems evolution,
but the point worth noting here is that $R_{cm}$ 
is a large number for a protostellar disk, whatever the ionization
source!
Ionization fractions far, far below unity can render
an astrophysical gas a near perfect electrical conductor.
It therefore makes a great deal of 
sense to begin by examining the behavior of an ideal
MHD fluid.

\section {Ideal MHD}

The fundamental equations of ideal (single-fluid)
MHD are mass conservation
(\ref{mass1})
\beq
{\dd\rho \over \dd t} + \del\bcdot (\rho\vv) = 0
\eeq
the dynamical equation of motion
\beq
{\dd\vv\over \dd t} +(\vv\bcdot\del)\vv = - {1\over\rho}\del
\left( P + {B^2\over 8 \pi}\right) -\del\Phi
+ ({\B\over 4\pi}\bcdot)\del\B
\eeq
where $\Phi$ is the (central) gravitational potential function,
and our newly simplified induction equation
\beq
{\dd\B\over \dd t} =\del\btimes(\vv\btimes \B )
\eeq
We shall work in a standard cylindrical coordinate
system $(R,\phi, z)$, where $R$ is the radius, $\phi$
the azimuthal angle, and $z$ the vertical coordinate.  
In these coordinates, the three components of the
equation of motion are
are
\beq
\left[ {\dd\ \over \dd t} +\vv\bcdot\del \right] v_R -{v^2_\phi\over R}
=
- {1\over\rho}{\dd\ \over \dd R}\left( P +{B^2\over 8 \pi} \right)
-{\dd\Phi\over \dd R}
+ {\B\over 4\pi\rho}\bcdot\del B_R -{B^2_\phi\over 4\pi \rho R},
\eeq
\beq\label{azi45}
\left[ {\dd\ \over \dd t} +\vv\bcdot\del \right](R v_\phi)
=
- {1\over\rho }{\dd\ \over \dd \phi}\left( P +{B^2\over 8 \pi} \right)
+ {\B\over 4\pi\rho}\bcdot\del (RB_\phi),
\eeq
\beq
\left[ {\dd\ \over \dd t} +\vv\bcdot\del \right] v_z
=
- {1\over\rho }{\dd\ \over \dd z}\left( P +{B^2\over 8 \pi} \right)
-{\dd\Phi\over \dd z}
+ {\B\over 4\pi\rho}\bcdot\del B_z ,
\eeq
and the three components of the induction equation are
\beq
\left[ {\dd\ \over \dd t} +\vv\bcdot\del \right] B_R =
-B_R \del\bcdot\vv +\B\bcdot\del v_R
\eeq
\beq
\left[ {\dd\ \over \dd t} +\vv\bcdot\del \right] {B_\phi\over R}
=
-{B_\phi\over R} \del\bcdot\vv +\B\bcdot\del \left(v_\phi\over R\right)
\eeq
\beq
\left[ {\dd\ \over \dd t} +\vv\bcdot\del \right] B_z =
-B_z \del\bcdot\vv +\B\bcdot\del v_z
\eeq
For many problems of interest, an important simplification
can be made to these equations.  The essence of rotational
dynamics is local.  Imagine going to a location $R_0$ in the
disk.  The Keplerian angular velocity $\Omega$ at $R=R_0$ is
$\Omega_0$.  We hold $\Omega_0$ fixed but allow $R_0$ to become
arbitrarily large.  Thus, $v_\phi (R_0)$ is unbounded,
but $v_\phi/R_0 =\Omega_0$ is finite.  

Next, we introduce local coordinates $(x, \alpha, z)$
defined by
\beq
R = R_0 +x, \qquad x\ll R_0,
\eeq
\beq
\alpha=\phi - \Omega_0t \ll  \pi
\eeq
and $z$ is unchanged.  We will formally introduce a new coordinate
$t'=t$.  This is desirable because we wish to take a time derivative
at constant $x,y,z$, not at constant $R, \phi, z$.  In fact,
\beq
{\dd\ \over\dd t'} = {\dd\ \over \dd t} + \Omega_0{\dd \over \dd \alpha}
\eeq
Partial derivatives with respect to $R$ and $\phi$ become partial
derivatives with respect to $x$ and $\alpha$ respectively.
The Lagrangian derivative
\beq
{\dd\ \over \dd t} +\vv\bcdot\del 
\eeq
becomes
\beq
{\dd\ \over \dd t'} +(\vv -R\Omega_0 \bb{e_\phi})\bcdot\del 
\eeq
(The $\del$ operator now formally involves $x$ and $\alpha$
derivatives.)  This transformation suggests that we
introduce a new azimuthal velocity,
\beq
w_\alpha = v_\phi -R\Omega_0
\eeq
and for notational
consistency we will use $w_x$ and $w_z$ for the radial
and vertical velocities, though they are identical to $v_R$ and $v_z$.

In the local approximation, we assume that the magnitude
$w$ is much smaller than the (formally infinite) rotation
velocity, and that $w$, the Alfven speed $v_A$, and
the thermal velocity $(P/\rho)^{1/2}$ are comparable
in magnitude.  
In the limit $R_0\rightarrow\infty$, the radial equation becomes
\beq\label{w1}
\left[{\dd\ \over \dd t'} +\bb{w}\bcdot\del\right] w_R
-2\Omega_0 w_\phi = R(\Omega^2_0-\Omega^2_K) 
- {1\over\rho}{\dd\ \over \dd x}\left( P +{B^2\over 8 \pi} \right)
+ {\B\over 4\pi\rho}\bcdot\del B_R,
\eeq
where the Keplerian angular velocity at $R$ satisfies
\beq
R\Omega^2_K = {\dd\Phi\over\dd R}
\eeq
Expanding to first order in $R$,
\beq
R(\Omega^2_0-\Omega^2_K)= -x{d\Omega^2_K\over dR} = 3\Omega^2_0
\eeq
The final form of the equation is
\beq
\left[{\dd\ \over \dd t} +\bb{w}\bcdot\del\right] w_x
-2\Omega w_y =  3\Omega^2 x                      
- {1\over\rho}{\dd\ \over \dd x}\left( P +{B^2\over 8 \pi} \right)
+ {\B\over 4\pi\rho}\bcdot\del B_x,
\eeq
where we have dropped the subscript $0$ and the prime $'$.
The vector components $x,y,z$
refer to the quasi-Cartesian system $dx=dR$, $dy=Rd\alpha$,
and $dz$ as before.  (See fig. 2.)  When non-Keplerian
disks are considered, $3\Omega^2$ should be replaced
by $d\Omega^2/d\ln R$.  

The transformed azimuthal ($y$) equation is
\beq\label{w1bis}
\left[{\dd\ \over \dd t} +\bb{w}\bcdot\del\right] w_y
-2\Omega w_x =  
- {1\over\rho}{\dd\ \over \dd y}\left( P +{B^2\over 8 \pi} \right)
+ {\B\over 4\pi\rho}\bcdot\del B_y,
\eeq
and the $z$ equation is
\beq
\left[{\dd\ \over \dd t} +\bb{w}\bcdot\del\right] w_z = 
-{\dd\Phi\over\dd z}
- {1\over\rho}{\dd\ \over \dd z}\left( P +{B^2\over 8 \pi} \right)
+ {\B\over 4\pi\rho}\bcdot\del B_z,
\eeq
The mass conservation and induction equations, like the $z$ 
equation of motion, show no change of form in these 
corotating coordinates:
\beq
{\dd\rho\over\dd t} + \del\bcdot (\rho\bb{w}) =0,
\eeq
\beq\label{w2}
{\dd\B\over
\dd t} = \del\btimes(\bb{w}\btimes\B )
\eeq
To summarize, we have transformed our general equations into
a coordinate frame rotating at the Keplerian
angular velocity $\Omega_K(R_0)$, restricting the spatial
domain to a small neighborhood around a particular fluid element.
The rotational dynamics appears in the form of a Coriolis force
$-2\bb{\Omega}\btimes\bb{w}$, and a centrifugal force that
cancels the main gravitational force,
leaving the residual tidal term  $3\Omega^2 x$.  These simplified
local equations, which are the magnetized version of what is
known as the {\em Hill system,} retain a rich dynamical content.
This includes the full development of MHD turbulence.

\subsection {Magnetorotational Instability}

Let us apply equations (\ref{w1})- (\ref{w2})
to the study of the paths of fluid
elements departing from the origin $x=y=z=0$.  These
{\em Lagrangian
displacements} are in the $xy$ orbital plane,
and depend only upon $z$.  We will ignore vertical
stratification here, so that the 
equilibrium is $z$ independent.  Thus, we may
assume a spatial dependence of $e^{ikz}$ 
for the perturbed fluid elements.  

We assume a very simple equilibrium
magnetic field: a constant
vertical component $B$, no components in the orbital
$xy$ plane.  If the displacement vector of an element of fluid in
the $xy$ plane is $\bb{\xi}$, then the perturbed magnetic
field at given location is
\beq
\delta\B = \del\btimes(\bb{\xi}\btimes\B)
\eeq
This result, which is true quite generally, is lengthy to prove
by a direct assault, 
but can be intuited rather easily.  The left side of the
exact equation (\ref{w2}) may be written $\delta\B/\delta t$,
the change in $\B$ at fixed location divided by the change in time.
In the equilibrium state this is zero.  Imagine perturbing this
pure state by giving each fluid an additional finite
velocity $\bb{U}$, but letting it act only for an infinitesimal
time $\delta t$.  Then
\beq
\delta\B = \del\btimes(\bb{w}\, \delta t \btimes\B)
+ \del\btimes(\bb{U}\, \delta t \btimes\B)= 0+
 \del\btimes(\bb{\xi} \btimes\B)
\eeq
where $\bb{\xi} =\bb{U}\delta t$ is the displacement of the fluid relative
to its equilibrium path.  This result
is perhaps more clear in integral form. If we integrate over a surface $A$
and use Stokes Theorem, there obtains
\beq
\int \delta\bb{B}\bcdot \bb{dA} = \int \bb{(\xi\times B)\cdot ds}
\equiv \int \bb{(ds\times \xi)\cdot B},
\eeq
where $\bb{ds}$ is a vector line element on the curve bounding $A$.
This states that whatever explicit change in magnetic flux there
is through $A$ as the surface is displaced by $\bb{\xi}$, it is 
precisely compensated by the
gain (or loss)
of magnetic flux through of the side of the cylinder 
(elemental area $\bb{ds\times \xi}$) swept out by $A$.  
This is just a round about way of saying that the magnetic flux 
through $A$ is conserved when moving {\em with} the fluid
itself.  

For the problem at hand,
\beq\label{btoxi}
\delta\B= \del\btimes(\bb{\xi} \btimes\B)=i\bb{k}\btimes(\bb{\xi} \btimes\B)
=i(\bb{k\cdot B})\bb{\xi},
\eeq
since $\bb{k\cdot \xi} =0$.  The magnetic term in equations
(\ref{w1}) and (\ref{w1bis}) is
\beq
{1\over4\pi\rho}\B\bcdot\del\delta\B = - {(\bb{k\cdot B})^2\over 4\pi\rho}
\bb{\xi}\equiv -(\bb{k\cdot  v_A })^2 \bb{\xi}
\eeq
We have introduced the so-called {\em Alfv\'en velocity}
\beq
\bb{v_A} = {\B\over \sqrt{4\pi\rho}}
\eeq
whose physical interpretation we discuss below.  For
the moment, note that the magnetic force is like tension or
a spring: it is always
restoring, and proportional to the displacement.  

In equations (\ref{w1}) and (\ref{w1bis}), the derivative
$$
{D\ \over Dt} \equiv {\dd \over \dd t} + \vv\bcdot\del
$$
is just the time derivative following a fluid element, so that
we will write $D\bb{w}/Dt ={\bb{\ddot\xi}}$, and of course 
$\bb{w}={\bb{\dot\xi}}$.  Finally, we will allow for the
possibility of
non-Keplerian rotation. 
The equations of motion
for the fluid element displacements $\xi_x$ and $\xi_y$ are then
\beq\label{xi1}
\ddot{\xi}_x -2\Omega\dot{\xi}_y =  - \left[ {d\Omega^2
\over d\ln R} + (kv_A)^2\right] \xi_x
\eeq
\beq\label{xi2}
\ddot{\xi}_y +2\Omega\dot{\xi}_x = 
- (kv_A)^2 \xi_y
\eeq

These are a simple set of coupled linear equations with time-dependent
solutions of the form $e^{i\omega t}$.  Without loss of generality,
we take $\omega>0$.  In the absence of rotation, 
the equations decouple completely and one finds
\beq
\omega = |kv_A|
\eeq
These are waves that travel along the magnetic field lines with
group and phase velocity $v_A$, exactly like waves on a string.
These so-called {\em Alfv\'en} waves are thus simple, nondipersive,
incompressible disturbances.  

On the other hand, in the absence of rotation, we find
\beq
\omega^2 = 4\Omega^2 + {d\Omega^2\over d\ln R} \equiv \kappa^2
\eeq
$\kappa^2$ is known as the epicyclic frequency.  In most
astrophysical applications, $\Omega$ decreases radially outwards
and $\kappa<4\Omega^2$.  The fluid dispacements for this mode execute
ellptical with a major to minor axis ratio of $\kappa/2\Omega$.  The major
axis lies along the azimuthal direction.  The sense of the rotation
around the ellipse is retrograde, opposite to the sense of $\Omega$.
These paths are known as epicycles.  Epicyclic motion corresponds
to the first
order departure from circular orbits for simple planetary motion.

The full dispersion relation resulting from equations (\ref{xi1})
and (\ref{xi2}) is more complicated than one might have expected:
\beq\label{simpdisp}
\omega^4 -[\kappa^2 +2(kv_A)^2]\omega^2 +(kv_A)^2\left[
(kv_A)^2 + {d\Omega^2\over d\ln R}\right] =0
\eeq
This is clearly not just a matter of adding Alfv\'enic and epicylic motion
in quadrature, something new is going on.  The dispersion relation is
a simple quadratic equation in $\omega^2$, and it is straightforward
to show that this quantity must be real.  Stability may then be investigated
by passing through the point $\omega^2\rightarrow 0$.  
Therefore, marginal instability corresponds to 
\beq\label{marginq}
(kv_A)^2 + {d\Omega^2\over d\ln R}=0
\eeq
and {\em instability} is present when
\beq
{d\Omega^2\over d\ln R} <0.
\eeq
Of course, this condition is nearly universally satisfied for
any type of astrophysical disk.  This is the simplest manifestation of
the {\em magnetorotational instability}, or MRI.

\subsection {Physical Interpretation}

The equations of motion (\ref{xi1}) and (\ref{xi2}) have a very simple
mechanical analogue: they are exactly the equations of two point masses
in orbit around a central body, bound together by a spring of frequency
$kv_A$ (Balbus \& Hawley 1992).

This mechanical analogy immediately offers a physical explanation of
the MRI (see figure [3]).  Two masses are connected by a spring, an inner mass $m_i$
and an outer mass $m_o$.  The mass $m_i$ orbits faster than the outer
mass $m_o$, causing the spring to stretch.  The spring pulls backwards
on $m_i$ and forwards on $m_o$.  The negative torque on $m_i$ causes it
to lose angular momentum and sink, while the positive torque on $m_o$
causes it to gain angular momentum and rise.  Thus, even though the
spring nominally supplies an attractive force, in a rotating frame this
drives the masses apart.  The continued stretching of the spring just
makes matters worse, and the process runs away.  The mass with the higher
angular momentum obtains yet more, the mass with less angular momentum
loses what little it has.  It is the ``same old story'' in a
dynamical context: the rich get richer and the poor get poorer.

\begin{figure}
\epsfig{file=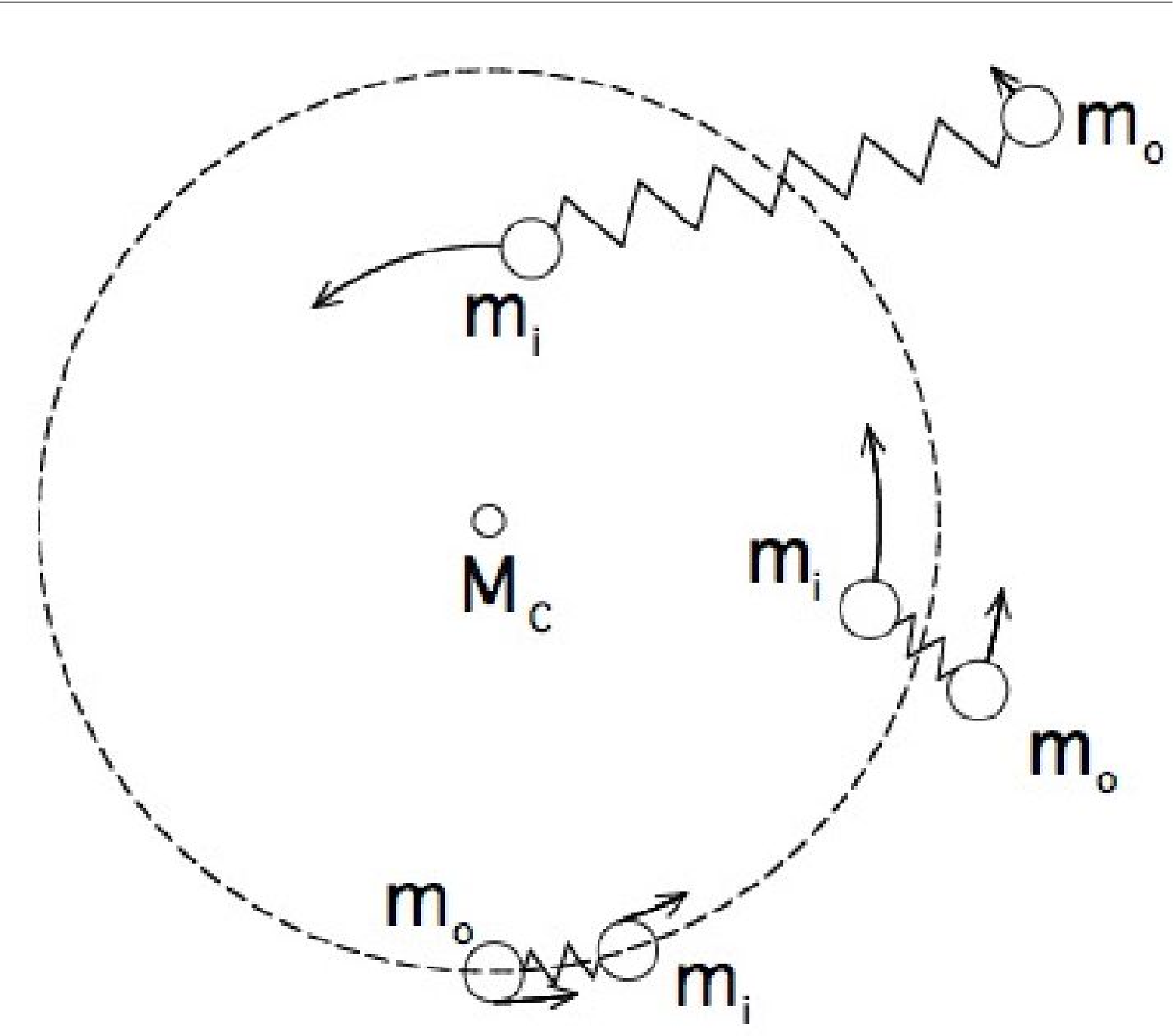, width = 10cm}
\caption {The magnetorotational instability.  Magnetic fields
in a disk bind fluid elements precisely as though they were masses
in orbit connected by a
spring.  The inner element $m_i$ orbits faster than the outer element
$m_o$, and the spring causes a net transfer of angular momentum
from $m_i$ to $m_o$.  This transfer is unstable, as described in the
text.  The inner mass continues to sink, whereas the outer mass
rises farther outward. ({\em Figure courtesy of H. Ji.)}}
\end{figure}

In a protostellar disk, it is the magnetic tension force that plays
the role of the spring, and the two masses are any two fluid elements
tethered by the field line.  The linear phase of the instability is the
exponentially growing
separation of the displaced fluid elements,  which is followed
by the nonlinear mixing of gas parcels from different regions of the disk.
The mixing seems to lead to something resembling a classical turbulent
cascade, though the details of this process, with different viscous and
resistive dissipation scales, remain to be fully understood.

Notice that angular momentum transport is not something that
happens as a consequence of the nonlinear development of the MRI,
it is the essence of the MRI even in its linear phase.  The very act
of transporting angular momentum from the inner to outer fluid elements
via a magnetic couple is a spontaneously unstable process.

\subsection {General Adiabatic Disturbances}

If $\Omega$ is a function only of cylindrical radius $R$, then 
for general magnetic field geometries,
local incompressible WKB disturbances with space-time dependence
\beq
\delta X \sim \exp (ik_R R +i k_z z -i\omega t)
\eeq
where $\delta X$ is the Eulerian perturbation of any flow
quantity, satisfy the following set of linearized dynamical equations:
\beq
k_R\delta v_R + k_z \delta v_z=0,
\eeq
\beq
-i\omega \delta v_R -2\Omega \delta v_\phi + i{k_R\over\rho}
\left[ \delta P + {\bb{B\cdot\delta B}\over 4\pi} \right]
-i{\bb{k\cdot B}\over 4\pi\rho}\delta B_R=0,
\eeq
\beq
-i\omega \delta v_\phi+{\kappa^2\over2\Omega} \delta v_R 
-i{\bb{k\cdot B}\over 4\pi\rho}\delta B_R=0,
\eeq
\beq
-i\omega \delta v_z  + i{k_z\over\rho}
\left[ \delta P + {\bb{B\cdot\delta B}\over 4\pi} \right]
-i{\bb{k\cdot B}\over 4\pi\rho}\delta B_z=0.
\eeq
The linearized induction equations are
\beq\label{indlin1}
-i\omega \delta B_R - i (\bb{k\cdot B})\delta v_R=0,
\eeq
\beq\label{indlin2}
-i\omega \delta B_\phi - \delta B_R{d\Omega\over
d\ln R} - i (\bb{k\cdot B})\delta v_\phi=0,
\eeq
\beq\label{indlin3}
-i\omega \delta B_z - i (\bb{k\cdot B})\delta v_z=0.
\eeq
Finally, the entropy equation is
\beq
i\omega \gamma {\delta\rho\over\rho} +\delta v_z{\dd\ln P\rho^{-\gamma}
\over \dd z} = 0,
\eeq
where $\gamma$ is the adiabatic index (not to be confused with the
collsion rate).
We have ignored background radial gradients in the entropy and pressure,
but retained their vertical gradients, in accordance with the assumption
of a thin disk.  

This set of linearized equations leads to the 
dispersion relation (Balbus \& Hawley 1991)
\beq
\left(k^2\over k_z^2\right)\tilde\omega^4 - \left[
\kappa^2 +\left(k_R\over k_z\right)^2 N^2\right]\tilde\omega^2
-4\Omega^2(\kva)^2=0.
\eeq
Here
\beq
\tilde\omega^2=\omega^2 -(\kva)^2, \quad k^2=k_z^2 +k_R^2,
\eeq
and 
\beq
N^2 = -{1\over\gamma\rho}{\dd P\over \dd z}
{\dd\ln P\rho^{-\gamma}\over \dd z}.
\eeq
$N$ is known as the \BV frequency. 
It is the natural frequency at which
a vertically displaced fluid element would oscillate in the disk due
to buoyancy forces.  In general there is also a contribution due
to radial gradients as well (Balbus \& Hawley 1991), but this usually may
be ignored in a rotationally supported (supersonic orbital speed) disk.  
Without any magnetic field, the dispersion relation becomes
\beq
\omega^2 = (k_z/k)^2 \kappa^2+(k_R/k)^2 N^2
\eeq
Since the displacement of the fluid element is incompressible,
the wave number ratio $k_z/k$ is a measure of the {\em radial}
displacement, whereas $k_R/k$ is a measure of the {\em vertical}
displacement.  There is no instability in this case, only
a wavelike response due to the restoring Corilois and buoyant
forces.  

Notice that $N$ vanishes in the disk plane by symmetry, it affects
only the behavior of disturbances at least one scale height or so
above $z=0$.  Moreover, the actual value of $N$ is likely to be
determined by radiative diffusion processes in the vertical
direction.  The radiation requires a source, which in disks
is the turbulence we are trying to explain in the first place!
(In protostellar disks, external heating of the upper layers is also
a source of departure from adiabatic behavior.)

The other limit of interest is $k_R=0$, which returns us to 
the dispersion relation of the previous section.  (The
expression $kv_A$ should of course be replaced by $\kva$.)
This is fortunate because it can be shown (Balbus \& Hawley 1992)
that the most rapidly growing modes are displacements in the
plane of the disk with $k_R=0$.  Since it is the differential
rotation that ultimately destabilizes, it is very sensible
that these displacements, which most effectively sample the differential
rotation, are the most unstable.  For a Keplerian rotation profile,
it may be directly computed from the dispersion relation
that the wave number of maximum growth is given by
\beq
\kva = (\sqrt{15}/4)\Omega\quad {\rm (wavenumber\ of\ maximum\ growth)},
\eeq
or $0.97\Omega$.  
By comparison, the largest unstable wavelength (i.e., smallest
unstable wave number) has a value of $1.73\Omega$ for $\kva$. 
The maximum growth rate is
\beq
|\omega| = (3/4)\Omega \quad {\rm (maximum\ growth\ rate)}
\eeq
This is a {\em very} large growth rate, with amplitudes growing
at a rate of $\exp(3\pi/2)\sim 111$ per orbit.

\subsection {Angular Momentum Transport}

The azimuthal equation of motion (\ref{azi45}) may be written
\beq\label{angmomcon}
{\dd(\rho R v_\phi)\over \dd t} +\del\bcdot\left[ R\left(\rho\vv v_\phi - \rho
\bb{v_A} v_{A\phi}
+ \left( P +{B^2\over 8\pi}\right)\bb{e_\phi}\right)\right]=0,
\eeq
where $v_{A\phi}$ is the azimuthal component of the Alfv\'en velocity.
This is an equation for angular momentum conservation, with
angular momentum density $\rho Rv_\phi$, and an azimuthally
averaged flux of
\beq\label{fj}
\bb{F_J} = \langle R\left(\rho\vv v_\phi - \rho
\bb{v_A} v_{A\phi}\right)\rangle
\eeq
In protostellar disks, we are interested in the radial transport
of angular momentum,
\beq
F_{JR} = R \rho (v_R v_\phi-v_{AR} v_{A\phi})
\eeq

It is most instructive to
begin with the simple case of linear instability in a uniform, vertical magnetic
field.  What is the lowest order flux that results
from these exponentially growing modes?
In equilibrium, $v_R$ and the Alfven velocities vanish.  The linear pertubation
of the radial velocity at a given spatial
location (a so-called Eulerian perturbation) will
be denoted $\delta v_R$.  Let the radial displacement $\xi_R$ of a fluid
element from its equilibrium location be given by
\beq
\xi_R = a e^{\gamma t} \cos(kz)
\eeq
where $\gamma$ and $k$ correspond to the maximum growth rate and its associated
wave number, and $a$ is a slowly varying amplitude. 
Then
\beq
\delta v_R = \dot{\xi}_R=\gamma \xi_R
\eeq
The azimuthal velocity $v_\phi$ consists of an unperturbed Keplerian
velocity $R\Omega$ plus a linear perturbation $\delta v_\phi$.  
The product of $\delta v_R$ and $\Omega$ contributes zero when a height
integration is performed because of the cosine factor, so we must
consider the direct product $\delta v_R\times\delta v_\phi$ to obtain
the lowest order contribution.  In calculating $\delta v_\phi$, note
that
\beq
\delta v_\phi = \dot{\xi}_\phi -\xi_R{d\Omega\over d\ln R}
\eeq
The subtraction is needed to eliminate the change in velocity a displaced
fluid element would make even if there were no physical change
in the rotation velocity at the new radial location: the actual change
in velocity $\delta v_\phi$ is due to the change in $\dot{\xi}_\phi$
the displaced fluid element makes {\em in excess} of $\xi_R d\Omega/d\ln R$.
Equation (\ref{xi1}) then gives
\beq
\delta v_\phi = \dot{\xi}_\phi -\xi_R {d\Omega\over d\ln R}
=\left( \gamma^2+k^2v_A^2\over 2\Omega \right) \xi_R
\eeq
and therefore the velocity contribution to the angular
momentum flux is
\beq
\langle \delta v_R \delta v_\phi\rangle = {\gamma\over 2\Omega}
(\gamma^2 +k^2v_A^2)\langle \xi_R^2\rangle
\eeq
Not surprisingly, it is positive (outward).  

To calculate the magnetic field correlation, we use equation
(\ref{btoxi}) to go between magnetic field fluctuations and
displacements.  Then, 
using (\ref{xi2}), we find
\beq
\xi_\phi = - {2\omega\gamma\over \gamma^2 +k^2v_A^2}\, \xi_R
\eeq
from which it follows, among other things, that $\dot{\xi}_\phi$ and
$\delta v_\phi$ have opposite signs.  Hence
\beq
{ \langle \delta B_R \delta B_\phi \rangle\over 4\pi\rho } =
- {2\Omega k^2v_A^2\gamma\over \gamma^2 +k^2v_A^2}\langle \xi_R^2\rangle
\eeq
Therefore
\beq
\langle \delta v_R \delta v_\phi-{\delta B_R \delta B_\phi\over
4\pi\rho}\rangle =
{\gamma\over 2\Omega}\left[ (\gamma^2+k^2v_A^2)^2 +4\Omega^2k^2v_A^2
\over \gamma^2 + k^2v_A^2\right]\langle \xi_R^2\rangle
\eeq
Using the dispersion relation (\ref{simpdisp}) to simplify this a bit, our expression
for the radial angular momentum flux is
\beq\label{fjr}
F_{JR} = \rho R \  {\gamma\over 2\Omega}\left[ {8\Omega^2k^2v_A^2\over
\gamma^2 +k^2 v_A^2} -\kappa^2 \right] \langle \xi_R^2\rangle
\eeq
Finally, integrating over height and defining the effective surface
density $\Sigma$ by 
\beq
\int \rho \langle \xi_R^2\rangle \, dz = \Sigma a^2 e^{2\gamma t}/2,
\eeq
the angular momentum flux for the most rapidly growing mode is found
to be
\beq
F_{JR}= {3R\Sigma \over 4} \Omega^2 a^2 e^{2\gamma t}
\eeq

This result bears some commentary.  First, the radial flux is always
positive for any unstable mode, not just the most rapidly growing.
This, as we shall see, is a direct consequence of energy conservation:
energy is extracted from the differential rotation to the fluctuations
only if the radial angular momentum flux is positive.  Next, note that
there is an outward angular momentum flux only to the extent that there
is a correlation in the velocity fluctuations.  This is also true in
turbulent flows, and everything that we have done in this section goes
through in much the same way if the fluctuations are turbulent as opposed
to wavelike.  The important physical point is that an instability that
leads to turbulence need not lead to enhanced angular momentum transport.
Only turbulence with strongly correlated velocity fields does this.
Strong correlations, in turn, are necessary to extract energy from
differential rotation.  This is indeed the free energy source of
shear-driven turbulence, but not any form of turbulence.  

This point was made in Stone \& Balbus (1996), in which the angular 
momentum transport resulting from {\em convective} instability was
studied.  The angular momentum was tiny in magnitude and {\em inwardly}
directed: it had the ``wrong'' sign.  Disk intabilities based on
adverse thermal gradients do not, in general, lead to systematically large
outward transfers of angular momentum.  When it comes to angular
momentum transport, not all turbulence is equivalent.

\subsection {Diffusion of a Scalar}

A classical problem in protostellar disk theory is to understand 
how dust particles are mixed with the gas.  The fact that the MRI
leads to vigorous radial angular momentum transport suggests that
other quantities may be transported as well.  In this section, we
will give an argument that shows that the radial diffusion of angular
momentum is indeed closely related to the radial transport of a 
conserved scalar quantity, say $Q$.  Assume that 
$Q$ is a fluid element label and satisfies an equation of the form
\beq
\left( {\dd\ \over \dd t} +\vv\bcdot\del\right) Q =0
\eeq
Then,
\beq
 {\dd Q \over \dd t} +\del\bcdot(\vv Q) - Q\del\bcdot\vv=0.
\eeq
Using mass conservation, this implies
\beq
{\dd Q \over \dd t} +\del\bcdot(\vv Q) + Q{D\ln \rho\over Dt}=0,
\eeq
where 
\beq
{D\ \over Dt} = \left( {\dd\ \over \dd t} +\vv\bcdot\del\right)
\eeq
Simplifying the equation for $Q$, we obtain
\beq
{\dd (\rho Q) \over \dd t} +\del\bcdot(\rho \vv Q)=0,
\eeq
which looks very much like the angular momentum conservation equation
(\ref{angmomcon}).  In a similar way, there will be a turbulent $Q$-flux given
by
\beq
F_Q = \rho \langle \delta\vv \delta Q \rangle
\eeq
Since $Q$ is conserved as we follow a fluid element, the so-called
{\em Lagrangian perturbation} $\Delta Q$, defined by
\beq
\Delta Q = \delta Q + \bb{\xi}\bcdot\del Q
\eeq
must vanish.  This is because $\Delta Q$ is constructed to
be the change in $Q$ following
a fluid element as it is displaced by a small distance $\bb{\xi}$,
and such changes {\em must} vanish since by definition $Q$ is does not
change along
fluid element paths.  Hence, our expression for the $i$th
component of $F_Q$ is
\beq
F_{Qi} = 
-\rho \langle \delta v_i \xi_j\rangle {\dd_j}Q
\eeq
This defines the diffusion tensor ${\cal D}_{ij}$,
\beq
{\cal D}_{ij}=\langle \delta v_i \xi_j\rangle
\eeq

In a height-integrated calculation, it is the $RR$ component
of ${\cal D}$ that is of importance.  In the linear regime
$\delta v_R=\gamma\xi_R$, so that 
\beq
{\cal D}_{RR}= \gamma\langle \xi_R^2\rangle
\eeq
In this case, equation (\ref{fjr}) gives a relationship
between the flux of angular momentum and the diffusion 
coefficient of a passive scalar,
\beq\label{fjrD}
F_{JR} =  {\rho R\over 2\Omega}\left[ {8\Omega^2k^2v_A^2\over
\gamma^2 +k^2 v_A^2} -\kappa^2 \right] {\cal D}_{RR}
\eeq
For the fastest growing linear mode, this gives $F_{JR}=2\Omega
{\cal D}_{RR}$.  

There is a sense in which the linear theory might find its way into
nonlinear turbulent diffusion.  In simulations the MRI appears to locally
stretch field lines and exponentiate velocity growth over a limited
duration of time, before one fluid element becomes mixed with another and
the process starts anew.  If such a picture is reasonably accurate, then
$F_{JR}\sim\Omega {\cal D}_{RR}$ may well be valid in turbulent flow.
(It is also, of course, what we might expect on the basis of simple
dimensional analysis alone!)  The important astrophysical point is that
if protostellar disks are MHD turbulent, they ought to be well-mixed.

\section {Energetics of MHD Turbulence}

\subsection {Hydrodynamic Considerations}

The quantity
\beq
\left(\rho\vv v_\phi - \rho
\bb{v_A} v_{A\phi}\right)
\eeq
plays two conceptually different roles
in the theory of accretion disks.  We have seen that it is 
intimately linked to the direct transport of angular momentum,
and in this section, we shall study in detail how it extracts free energy
from the large scale differential rotation.  

Let us begin with the relatively simple case of an adiabatic
nonmagnetized gas.  The azimuthal velocity is decomposed into
a time-steady large rotational component $R\Omega$ plus
a fluctuating component $u_\phi$:
\beq
v_\phi = R\Omega +u\phi
\eeq
we will assume neither than $u\phi$ is small compared with
$R\Omega$ nor that the mean value of $u_\phi$ vanishes, though
in practice both might well be the case.  We will write $u_R$
and $u_z$ for the radial and vertical velocity components respectively,
and $\vv$ for the full velocity vector.
Once again, we think of these as fluctuations, but our treatment
will in fact be exact.  

The adiabatic radial equation of motion is
\beq
\rho\left[ {\dd\ \over \dd t} 
+\bb{v\cdot\nabla}\right] u_R-{\rho\over R}(R\Omega+u_\phi)^2 = 
-{\dd P\over \dd R} - \rho {\dd\Phi\over \dd R}
\eeq
Multiplying by $u_R$ and regrouping:
\beq
\rho\left[ {\dd\ \over \dd t} 
+\bb{v\cdot\nabla}\right] \left(u_R^2\over2\right)-2\rho\Omega u_\phi u_R -
\rho{u_Ru_\phi^2\over R} = -u_R\left({\dd P\over \dd R}+\rho{\dd\Phi_{eff}
\over \dd R} \right)
\eeq
where
\beq
\Phi_{eff} = \Phi -\int^R  s\Omega^2(s) \, ds
\eeq
Exactly the same manipulations with the $\phi$ and $z$ equations
produce 
\beq
\rho\left[ {\dd\ \over \dd t}
+\bb{v\cdot\nabla}\right] \left(u_\phi^2\over2\right)+\rho u_\phi u_R
{\kappa^2\over 2\Omega}+
\rho{u_Ru_\phi^2\over R} = -u_\phi{\dd P\over R\dd \phi}
\eeq
\beq
\rho\left[ {\dd\ \over \dd t}
+\bb{v\cdot\nabla}\right] \left(u_z^2\over2\right)
= -u_z\left({\dd P\over \dd z}+\rho{\dd\Phi_{eff}
\over \dd z} \right)
\eeq
where we have assumed that $\Phi$ is independent of $\phi$.  
Adding the three dynamical equations and using mass conservation
leads to 
\beq\label{enu1}
{\dd\ \over \dd t}\left( {1\over 2}\rho u^2\right)
+\del\bcdot(\rho u^2\vv/2) +\rho u_R u_\phi {d\Omega\over d\ln R}
 =-\bb{u\cdot\nabla}P -\rho\bb{u\cdot\nabla}\Phi_{eff}
\eeq
where $u^2=u_R^2+u_\phi^2+u_z^2$.  
Yet another use of mass conservation and a regouping of the
pressure term gives us
\beq
{\dd\ \over \dd t}\left( {1\over 2}\rho u^2 +\rho\Phi_{eff}\right)
+
\del\bcdot\left( \vv({1\over2}\rho u^2 +\rho\Phi_{eff}) +P\bb{u} \right)
=P\del\bcdot\bb{u} -\rho u_R u_\phi {d\Omega\over d\ln R}
\eeq
We have already at hand the main structure of our energy equation.
The left side is in conservation form with a well defined energy density
and energy flux, with the fluctuations isolated.  (Notice that the
azimuthal average of the energy equation, which we ultimately will
be working with, has no rotational terms in the energy flux divergence.)  
Sources of energy fluctuations are work done by pressure (which
may cause heating or cooling), and the all
important $R\phi$ stress coupling to the differential rotation.  

\subsection {The Effects of Magnetic Fields}
When magnetic fields are included, everything goes through as before,
except that the right side of equation (\ref{enu1}) contains the terms
\beq\label{magen1}
-\bb{u\cdot\nabla}P_{tot} -\rho\bb{u\cdot\nabla}\Phi_{eff}
+{\bb{u}\bcdot[\bb{B\cdot\nabla}]\bb{B} \over 4\pi}
\eeq
The final magnetic term in the above may be written in an index
notation as
\beq
{u_j B_i\dd_i B_j\over 4\pi}
\eeq
where $i,j,k$ take the values $x,y,z$, and repeated indices are
summed over.  $\dd_i$ denotes the the partial derivative with
respect to the $i$th spatial variable.  

To make further progress, we need the induction equation:
\beq
\left({\dd\ \over \dd t} + \vv\bcdot\del\right)\bb{B} = 
-\bb{B}\del\bcdot\vv +(\bb{B}\bcdot\del)\vv.
\eeq
Take the dot product of this with $\bb{B}$ and write the last
term in component form:
\beq\label{indB1}
\left({\dd\ \over \dd t} + \vv\bcdot\del\right){B^2\over2}
=-B^2\del\bcdot\vv+ B_jB_i\dd_i v_j
\eeq
Now,
\beq
B_jB_i\dd_i v_j
=\dd_i(B_iB_j v_j) -v_j\dd_i(B_jB_i) =\del\bcdot(\vv\bcdot\bb{B}\bb{B}) -R\Omega
[\bb{B}\bcdot\del\bb{B}]_\phi - u_jB_i\dd_iB_j
\eeq
where we have used $\dd_iB_i=0$.  
Note that with the last term, we make contact with the energy equation terms
(\ref{magen1}).  The subscript $\phi$ on the penultimate term denotes
a vector component:
\beq
R\Omega [\bb{B}\bcdot\del\bb{B}]_\phi = \Omega[\bb{B}\bcdot\del(RB_\phi)]
=\del\bcdot(R\Omega\bb{B}B_\phi)- B_\phi B_R
{d\Omega\over d\ln R}
\eeq
Therefore, the right side of equation (\ref{indB1}) becomes
\beq
-B^2\del\bcdot\vv+\del\bcdot[(\bb{u}\bcdot\bb{B})\bb{B}] +B_\phi B_R
{d\Omega\over d\ln R}- u_jB_i\dd_iB_j
\eeq
Combining this result with the left side of the equation
(\ref{indB1}) then, after some
cancellation, given us an expression for $u_jB_i\dd_iB_j$:
of terms,
$$
- u_jB_i\dd_iB_j= \left({\dd\over \dd t} +\Omega{\dd\ \over \dd\phi}\right){B^2\over 2}
+\del\bcdot\left( \bb{u}{B^2\over 2} - (\bb{u\cdot B})\bb{B} \right)\qquad {\ }
$$
\beq
{\ }\qquad\qquad {\ }+{B^2\over2} \del\bcdot\bb{u} -B_\phi B_R {d\Omega\over d\ln R}
\eeq
Armed with this result, we return to (\ref{magen1}) and make a substitution
for the final term.  The resulting energy equation can be simplified
to
$$
{\dd\ \over \dd t}\left( {1\over2}\rho u^2 +\rho\Phi_{eff} +{B^2\over 8\pi}\right)
+\del\bcdot\left[ \vv \left({1\over2}\rho u^2+\rho\Phi_{eff}\right) +P\bb{u}
+{\bb{B\times({u}\times B )}\over 4\pi}\right]+
$$
\beq\label{magen2}
+\del\bcdot\left(\bb{e_\phi} R\Omega {B^2\over 8\pi}   \right) = P\del\bcdot\bb{u}
-\rho(u_R u_\phi - v_{AR}v_{A\phi}){d\Omega\over d\ln R}
\eeq
The pressure term on the right side of the above equation can be eliminated using
the thermal energy equation
\beq\label{thermen}
{3\rho\over2} \left( {\dd\ \over\dd t} + \vv\bcdot\del \right) {P\over\rho} =
- P \del\bcdot\bb{u} -\rho{\cal L}
\eeq
(since $\del\bcdot\vv=\del\bcdot\bb{u}$).  We have introduced the
radiative energy loss term per unit volume $\rho{\cal L}$. 
Carrying through the elimination of the pressure term and averaging over
$\phi$ lead to the total energy equation:
\beq\label{magfinal}
{\dd{\cal E} \over \dd t}+\del\bcdot{\bb{\cal F}}=
-\rho(u_R u_\phi - v_{AR}v_{A\phi}){d\Omega\over d\ln R}
- \rho {\cal L} \qquad {\rm (\phi\ averaged)}
\eeq
where ${\cal E}$ is the energy density in fluctuations,
\beq
{\cal E} =  {1\over2}\rho u^2 +\rho\Phi_{eff} +{B^2\over 8\pi}
+{3P\over 2}, 
\eeq
and ${\bb{\cal F}}$ is the corresponding flux,
\beq
{\bb{\cal F}}= \bb{u} \left({1\over2}\rho u^2+\rho\Phi_{eff}\right) +{5
P\over 2}\bb{u}
+{\bb{B\times({u}\times B )}\over 4\pi}
\eeq

We have completely ignored dissipation effects (viscosity and resistivity).
What effect would this have on our final equation (\ref{magfinal})?  The answer
is essentially none.  While it is true that new dissipation terms would
appear on the right side of equation (\ref{magen2}), they would also appear
with the opposite sign in equation (\ref{thermen}).   They would completely
cancel on the right side of
the final equation (\ref{magfinal}): dissipation is not a loss
of energy, but a conversion of mechanical or magnetic field energy
into heat.  Additional small energy flux terms
{\em would} appear (within the divergence operator)
in connection viscosity and resistivity, but these are generally
negligible compared with the dynamical terms.  The essential physical point
is that total energy is conserved in the presence of dissipation even if mechanical
energy is lost.  Only radiative processes can remove energy from the disk.

We interpret equation (\ref{magfinal}) as saying that energy is exchanged
with the differential rotation 
at a volumetric rate $- T_{R\phi}d\Omega/d\ln R$, where
\beq
T_{R\phi} = \rho(u_R u_\phi - v_{AR}v_{A\phi})
\eeq
is the dominant component of the 
stress tensor.  The energy made available from the differential rotation
may in principle remain in velocity fluctuations in the form of waves, but
in a thin, cool disk it is more likely that this energy will be locally
dissipated, and subsequently radiated.  This is because 
if the energy flux varies radially over a scale of order
$R$ itself, then the $T_{R\phi}$ source term on the right
will be larger than any term on the left (by an amount of order
$u/(R\Omega)$).  The stress can only be counterbalanced by
the radiative loss term:
\beq\label{enbal}
- T_{R\phi}d\Omega/d\ln R =\rho {\cal L}
\eeq
Therefore, although $- T_{R\phi}d\Omega/d\ln R$ is itself a nondissipative
source term, in these so-called local models, it works out to be the rate
at which energy is dissipated (and ultimately lost by radiation) as well.
Bear in mind that the identification of the stress term with dissipation
follows from a reasonable assumption about the behavior of thin disks:
local dissipation of the fluctuations.  Dissipation is in no way a
fundamental property of the left side of equation (\ref{enbal}), which
under different conditions could
just as well be an energy source or reserve for purely adiabatic waves.
There are many papers in which this point is misunderstood: {\em caveat
lector.} 

\section {Resistive and Hall terms}

\subsection {Local Dispersion Relation}

Extensive regions in real protostellar disks are in a regime far from ideal MHD, 
in which Ohmic decay and Hall electromotive forces are important
(e.g. figure 1).  It is important to understand how the MRI behaves
under these conditions.

The fastest growing modes, as usual, correspond to axial wavenumbers
$\bb{k}= k\bb{e_z}$.  Fortunately, these are also the simplest to
treat analytically, and we will limit our discussion to this class
of disturbance.

The linearly perturbed Hall term in the induction equation is to leading
WKB order
\beq
- \del\btimes \left[{\bb{\delta J\times B}\over e n_e}
\right]=\del\btimes\left[{c\bb{B}\over 4\pi en_e}
\btimes (i\bb{k}\btimes \bb{\delta B})\right],
\eeq
since $\bb{\delta J}=(c/4\pi)(i\bb{k\times\delta B})$.
Taking the curl and simplifying allows us to write the
right side as 
\beq
{c\over 4\pi e n_e}(\bb{k\cdot B})(\bb{k\times\delta B}).
\eeq
(In a fully ionized disk, the presence of $n_e$ in the denominator
renders these terms negligbily small.)
For our particular problem, $\bb{k}$ is axial and $\bb{\delta B}$
lies in the disk plane, and the above reduces to
\beq
{ck^2 B \over 4\pi e n_e}(\delta B_R\bb{e_\phi} - \delta B_\phi\bb{e_R}).
\eeq
Including the effects of resistivity, the linearized induction equations 
(\ref{indlin1}-\ref{indlin2}) become: 
\beq\label{indlin4}
(-i\omega +\eta k^2)\delta B_R +{ck^2 B_z\over4\pi e n_e}\, \delta B_\phi
- i (\bb{k\cdot B})\delta v_R=0,
\eeq
\beq\label{indlin5}
(-i\omega +\eta k^2) \delta B_\phi - \delta B_R\left(
{d\Omega\over d\ln R} +{ck^2 B_z\over4\pi e n_e}\right)
- i (\bb{k\cdot B})\delta v_R=0,
\eeq
and the $z$ equation is not needed (trivially satisfied).  
The linearized dynamical and energy equations from section (4.3) 
remain unchanged.

The Hall electromotive force terms introduce a phase shift in the induction
that leads to a circularly polarized component of what would otherwise
be an Alfv\'en or slow mode.  When an instability is present due to
differential rotation, this phase shift can either destabilize or stabilize
depending upon the {\em sign} of $B_z$.  

To see this more quantitatively, we must study the dispersion relation
that follows from our linearized set of equations.  To this end,
we define a Hall parameter $v_H$ which has the dimensions of a
velocity (Balbus \& Terquem 2001),
\beq
v_H^2 = {\Omega B_z c \over 2\pi en_e}.
\eeq
Notice that if we take the sense of rotation to be positive, that 
$v_H^2$ can have either sign depending upon whether $B_z$ is
positive or negative.  
With $\sigma=-i\omega$, our dispersion relation is
\beq\label{halldisp}
\sigma^4 + 2 \eta k^2 +{\cal C}_2 \sigma^2 + 2\eta k^2(\kappa^2
+k^2v_A^2)\sigma +{\cal C}_0 = 0,
\eeq
where the constants ${\cal C}_2$ and ${\cal C}_0$ are given by
\beq\label{calc2}
{\cal C}_2 = \kappa^2 +2 k^2 v_A^2 +\eta^2 k^4  +
{k^2v_H^2\over 4\Omega^2} \left( {d\Omega^2\over d\ln R} +k^2
v_H^2\right)
\eeq
\beq\label{calc0}
{\cal C}_0 = k^2\left( v_A^2 +{\kappa^2 v_H^2\over 4\Omega^2}\right)
\left( {d\Omega^2\over d\ln R} + k^2v_A^2+k^2v_H^2\right)
+\kappa^2\eta^2 k^4
\eeq

\subsection {Stability}

A sufficient condition for instability is that ${\cal C}_0 < 0$.  
When compared with the purely Ohmic requirement
\beq
k^2v_A^2 \left( k^2v_A^2 +{d\Omega^2\over d\ln R} \right)
+\kappa^2\eta^2 k^4 <0
\eeq
The Hall condition ${\cal C}_0 <0$ is much more easily satisfied, since
$v_H^2$ can change sign.  
Perhaps most striking is the possibility that the intuitive
result that large wavenumbers are stabilized
can be lost when $v_H^2<0$, and that {\em all} wavenumbers can be
unstable, even in the presence of finite
resistivity.  (See figure [4].)
This may be shown straightforwardly from the definition
of ${\cal C}_0$; the reader may wish to consult Balbus 
\& Terquem (2001) for explicit details.  In addition to requiring
counteraligned magnetic and rotation axes, this state also
demands 
a magnitude ratio of $v_A^2$ to $v_H^2$ lies between about 2 and 4  
(for a Keplerian disk).  When the field is aligned with the rotation axis,
the range of unstable wavenumbers is more restricted than that found for 
the ideal-MHD MRI.  

\begin{figure}
\epsfig{file=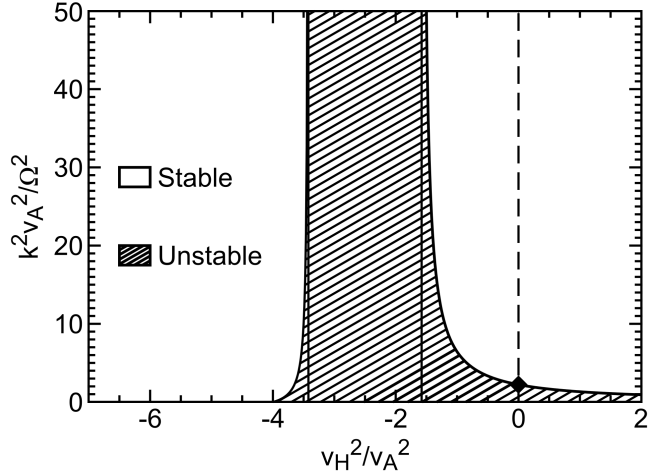, width = 10cm}
\caption{Range of unstable wavenumbers for a representative disk
in the Hall-Ohm regime.  The black diamond corresponds to the
MRI without the Hall effect.  The resistivity is chosen
to be $\kappa^2\eta^2/v_A^4=0.35$.
From Balbus \& Terquem (2001).}
\end{figure}

\subsection {Numerical Simulations of the Hall--Ohm--MRI.}

What does the preceding section imply for the MHD transport properties
of protostellar disks?  The first local simulatons of the nonlinear
development of the MRI in the presence of Ohmic resistivity are those
of Fleming, Stone, \& Hawley (2000).  These authors found the important
result that the nonlinear turbulent state is easier to maintain when a
mean vertical magnetic field is present.  This is sometimes misunderstood
as a requirement that a disk must have a global vertical magnetic field to
maintain MRI turbulence, but in fact all that the very local calculations
of Fleming et al. (2000) really tell us is that on this scale, working
with a precisely zero mean field is probably too restrictive an
approximation for a real disk.  This study found a difference of two
orders of magnitude in the critical magnetic Reynolds number (defined
here as the product of the isothermal times the box size divided by
the resistivity) needed to sustain turbulence, depending upon whether
a vertical field was present or not, the vertical field runs being the
easier to sustain.  It is now understood that very high resolution grids
are needed for the local study of the MRI in a shearing box with zero
mean field, so these early results should not be used quantitatively.
With this precaution understood, it is interesting to note that for the
vertical field runs, the Fleming et al.\ finding seems to be not very
different from the ${\mathbb {L}}=1$ criterion discussed earlier.

This nature of the nonlinear behavior of a Hall fluid was examined in a
numerical investigation by Sano \& Stone (2002a,b).  These authors carried
out an extensive study of the local properties of Hall-MRI turbulence
in the shearing box formalism.  The results are somewhat surprising.

In linear theory, counteraligned magnetic and rotation axes produce
a very broad wavenumber spectrum of instability, and should be more
unstable than the aligned case.  What Sano and Stone found, however,
was just the opposite: the aligned case showed greater levels of field
coherence and higher rms fluctuations than the counteraligned case.
The reason for this is interesting, and illustrative of the hazards of
extrapolating directly from linear theory.

In the case of counteraligned axes, the extended wavenumber of spectrum
of linear MRI instability results, in its nonlinear resolution, in a
highly efficient turbulent cascade to smaller and smaller scales.  In a
finite difference numerical simulation, this cascade terminates at the
grid scale where energy is ultimately lost.  By contrast, the aligned
case shows instability only at longer wavelengths (small wavenumbers).
At larger wavenumbers, the response of the fluid is wavelike, and this
is likely to make a local small scale turbulent cascade considerably
less efficient.  The result: a less lossy system and greater large scale
field coherence.

Finally, when simulations were done with half the field aligned
with the rotation axis and half counteraligned, the Hall effect
tended to wash out, and the results were similar to resistive
MHD turbulence without the additional electromotive forces.  The
criterion for the onset of turbulence depended only the relative
strength of the resistivity,
as measured by the dimensionless number ${v_A^2/\eta\Omega}$.
This so-called {\em Lundquist number} must be greater than
unity or MRI-induced turbulence is not maintained, whether
Hall electromotive forces are present or not, according to
Sano \& Stone.  The presence of the Hall terms in the induction
equation do, however, promote a slightly elevated rms level 
when turbulence can be maintained.

The Hall numerical studies are fascinating and suggestive, but caution
is necessary before taking the final leap from simulation to real disks.
Note, for example, the role played by the numerical grid scale losses in
our discussion of the nonlinear resolution of the counteraligned case.
Nature works without a grid, however, and it is not clear what scale
ultimately intervenes at high wavenumbers in Hall protostellar disks.
Moreover, it is now understood that there
are important resolution questions for local shearing box simulations
with no mean field, that have only recently been brought to light
(Fromang et al. 2007; Lesur \& Longaretti 2007).
The important pioneering studies of Sano \& Stone (2002a,b) merit a
thoroughgoing follow-up on the much larger numerical grids that are
currently available to numericists.

\section {Alpha models of protostellar disks}

\subsection{Basic principles}

We are now in a position to assemble a basic ``$\alpha$ model'' of a
protostellar disk.  An important assumption in such models
is that angular momentum and mechanical
energy are transported radially through the disk.  We therefore work with
height and azimuthally integrated quantities.  Under these conditions,
Balbus \& Papaloizou (1999) have shown how the fundamental equations
of MHD lead directly to the $\alpha$ formalism.  

The velocities consist of a mean part plus a fluctuation of zero mean.
For the azimuthal velocity $v\Phi$,
the mean will be taken to be the Keplerian
velocity
\beq
(R\Omega)^2 = {GM\over R}
\eeq
where $M$ is the central mass and $G$ is the Newtonian constant.  For the
radial velocity $v_R$, the mean motion corresponds to the very slow
inward accretion drift, $v_2$.  The azimuthal fluctuation velocity
$u_\phi$ is much less
than $R\Omega$, whereas the radial fluctuation velocity $u_R$ is much larger
than the inward drift.  These claims will shortly be quantified.  
The magnetic field Alfv\'en velocities are assumed to be of the
same order as the $u$ velocities, namely less than or of order the
isothermal thermal sound speed $c_S$.   To summarize,
\beq
v_2 \ll u, v_A, c_S \ll R\Omega.
\eeq

Under steady-state conditions, the local disk accretion
rate has a time averaged value of 
\beq
\mu\equiv 2\pi R \langle \rho v_2 + \delta\rho\, \delta u_R\rangle
\eeq
The height integrated form of this is $-\dot{M}$, the mass accretion
rate:
\beq
\int \mu\,  dz = -\dot M,
\eeq
defined so that $\dot M$ is a positive constant.  A useful average of $v_2$
is the drift velocity defined by
\beq
v_d \equiv - {\dot{M}\over 2\pi R \Sigma}
\eeq
where $\Sigma$ is the disk surface density.  

In steady-state, the height-integrated average radial
angular momentum flux is proportional to $1/R$.  In other words,
\beq\label{r1}
- \dot{M} R^2\Omega + 2\pi \Sigma R W_{R\phi} = C/R
\eeq
where $W_{R\Phi}$ is defined by
\beq
\int \langle \rho u_R u_\phi - \rho v_{AR} v_{A\phi}\rangle\, dz
=\Sigma W_{R\phi},
\eeq
and $C$ is a constant to be determined.  Traditionally, this has
been done by asserting that $W_{R\phi}$ is proportional to 
the radial gradient of $\Omega$, and at some point before the surface of
the star is reached this must vanish.  $C$ is then determined at
this point (Pringle 1981).  

This approach now seems dated, and especially inappropriate for a
protostellar disk in which the magnetic interactions between the disk
and star may become quite complex at small radii.  Instead, let us note
that at small $R$, $C/R$ must be very small if $\Sigma W_{R\phi}$ does
not blow up, which seems quite reasonable.  If $C/R$ is very small at the
inner edge of the disk, it is clearly negligible in the body of the disk,
and we shall set $C=0$ with the understanding that our solution should
not be taken to the stellar surface.  Then, (\ref{r1}) reduces to
\beq
v_d = - {W_{R\phi}/R\Omega}
\eeq
showing that $v_d$ (and $v_2$ which is of the same order) is
very small compared with the $u$ velocities.  Alternatively,
\beq\label{dub}
\Sigma W_{R\phi} = {\dot M} R \Omega /2\pi.
\eeq
This is a particularly useful result because $W_{R\phi}$
appears in the energy balance equation (\ref{enbal}):
\beq\label{dub1}
-\Sigma W_{R\phi}{d\Omega\over d\ln R}
=- {d\Omega\over d\ln R} \int T_{R\phi}\, dz = \int \rho {\cal L}\, dz
= 2 \sigma T_{eff}^4.
\eeq
In the last equality, we have equated the radiated energy per unit surface
of the disk to twice the blackbody emissivity, the factor of two represents
two radiating surfaces.  Here $\sigma$ is the Stefan-Boltzmann constant
and $T_{eff}$ is the effective blackbody surface temperature.  Combining
(\ref{dub}) and (\ref{dub1}) then gives us
\beq\label{ts}
T_{eff}^4  = {3GM\dot{M}\over 8\sigma \pi R^3},
\eeq
relating the potentially observable disk surface temperature to
the central mass, accretion rate, and radial location.  The unknown
turbulent stress parameter $W_{R\phi}$ has conveniently vanished!

Equation (\ref{ts}) is in a restricted sense ``exact.''
It is based on the assumption of time-steady conditions and thermal
radiation, but it is independent of the explicit nature of the turbulence,
as long as it is {\em local}.  If these conditions are all met, 
equation (\ref{ts}) is a simple matter of energy conservation.  
To go beyond this result, something has to be said directly about
$W_{R\phi}$.  In $\alpha$ disk theory, that condition is
\beq\label{alpha1}
W_{R\phi} = \alpha c_S^2
\eeq
where $\alpha$ is assumed to be constant, but otherwise unknown.
In contrast to the simple and plausible assumption that turbulence
is locally dissipated, this ``$\alpha$ assumption'' is far from obvious,
and in a strict mathematical sense almost certainly wrong.

The original justification for this form of the stress tensor was
based on the notion of hydrodynamical turbulence and the idea that
the velocity fluctuations would be restricted to some fraction of the 
sound speed (fluctuations in excess would cause shocks).  Shakura \& Sunyaev
(1973) consider magnetic stresses, however, and argue that they fit
within the $\alpha$ formalism as well since Alfv\'en velocities in 
excess of $c_S$ are also dynamically unlikely.  

The real problem with the prescription (\ref{alpha1}) is that turbulence
is just too complicated.   Not only are long term averages very hard to
define (a problem even for a result like equation [\ref{ts}]), it is also
entirely possible that $\alpha$ could vary in a complex nonsystematic way
by an order of magnitude or more from one part of the disk to another.
The primary justification for (\ref{ts}) is that many scaling results
are very insensitive to $\alpha$.  On dimensional grounds $c_S$ is is
certainly an important local characteristic velocity, but it is not yet
clear under what conditions a ``background'' magnetic field might also be
providing a mean Alfv\'en velocity that limits or guides the turbulence.

Continuing with our $\alpha$ model, to link the surface temperature $T_{eff}$
with the midplane temperature $T$ (used in quantities like $c_S$) we need
to introduce explicit vertical structure into the problem.  The hydrostatic
equilibrium equation is
\beq
-{\dd P\over\dd z} = {GM\rho z\over R^3} = \rho z\Omega^2
\eeq
This may be solved in conjunction with a detailed energy equation,
but it seems best for illustrative purposes to note that this equation
serves to provide a decent and very simple estimate of the disk half
thickness, $H=c_S/\Omega$.  The energy equation for simple vertical radiative
diffusion defines the local radiative energy flux $F_\gamma$ as:
\beq
F_\gamma = {4\sigma\over 3}{dT^4\over d\tau}.
\eeq
The ``optical depth'' $\tau$ is given by
\beq
d\tau = - \rho\kappa dz
\eeq
where $\kappa$ is the opacity of the disk (units: cross sectional area per
unit mass).  In the simplest possible model, $F_\gamma$ is a constant 
given by $\sigma T_{eff}^4$, and the temperature at the midplane $T$ is
then
\beq
T^4={3\tau T_{eff}^4\over 4}
\eeq
where $\tau$ is the optical depth from the outer disk surface to the midplane.
(In this simplest of all possible models, we set $\tau=\kappa H$, where $\kappa$
is evaluated at the midplane temperature.)

The surface temperature of a disk with a mass accretion rate
of $10^{-8} M_\odot$ per year around a $1 M_\odot$ star is 
$85R_{AU}^{-0.75}$K, where $R_{AU}$ is the radius in astronomical
unit.   The midplane temperature $T$ is a factor $\sim \tau^{1/4}$,
larger, typically a factor of 5 or so larger.  At what value of
$T$ would we expect the ionization fraction to reach the critical
value of $10^{-13}$ we found earlier, corresponding to a Lundquist
number of unity?  Put in somewhat different terms, 
is our model of self-sustaining
MHD turbulence self-consistent?

To answer this, we need to address the physics of thermal ionization.
In the low ionization regime we are working, thermal electrons are supplied
by trace alkali elements, notably potassium, with an ionization potential
of only 4.341 eV.  Even this modest value corresponds to an effective temperature
of 50,375 K, well above the range of $10^2$ to $10^3$K we expects, and
potassium will barely be ionized.  

The equation governing the ionization fraction $x$ is the
Saha equation\footnote{The discussion presented here 
presumes that the ionization
reaction rates are sufficiently rapid to keep up with the dynamical
time scales of any turbulence present.  At low ionization fractions
this breaks down, and a single temperature may not be enough to 
characterize the ionization state (Pneuman \& Mitchell 1965).}.
In this barely ionized regime, it may be written (Stone et al. 2000)
\beq
x^2 = as\left(2.4\times 10^{15}\over n \right) T^{3/2} \exp(-50375/T)
\eeq
where $a$ is the abundance of potassium ($~10^{-7}$), $n$ is the ratio
of the dominant $H_2$ molecules, and $s$ is a ratio
of statistical weights, expected to be near unity.  We may rewrite this
equation as
\beq
T =  - {50375\over \ln X}
\eeq
where 
\beq
X= x^2T^{-3/2} (n/2.4\times 10^{15})
\eeq
With $x=10^{-13}$ and the final density factor anything between
0.01 and 1 ($0.1 M_\odot$ of gas spread out over a region of 
10 AU yields a density of about $5\times 10^{13}$ per cc)
gives values for $T$ close to $10^3$K, which we will adopt as
a working number.  

Therefore, a surface temperature of about $200$K is required to
attain a midplane temperature of $10^3$, and thereby ensure
a critical level of
ionization.  This is the key issue for MHD theories of
protostellar disks: {\em beyond $0.2-0.3$AU, or $\sim 3\times 10^{12}$cm,
the heat generated by the dissipation associated with local turbulence
is insufficient to keep the the disk magnetically well-coupled.}
Where and how do protostellar disks maintain good magnetic coupling?

\section {Ionization Models of Protostellar Disks}
\subsection {Layered accretion}

The above considerations indicate that dissipative heating from MHD
turbulence self-consistently generates enough heat to maintain the
minimum thermal levels of requisite ionization only with a few 0.1 AU.
Nonthermal sources of ionization are therefore of great interest to
protostellar disk theorists.  The principle ionizing agents that have
been studied are cosmic rays, X-rays, and radioactivity.  

Gammie (1996) argued that the low density extended vertical layers of a
protostellar disk would be exposed to an ionizing flux of interstellar
cosmic rays.  Just as in models of molecular cloud ionization, cosmic rays
would maintain a minimal level of ionization in the upper disk layers.
The range of the low energy galactic cosmic rays is about 100 g/cm$^2$.
This is much less than the disk column density at $\sim 1$ AU in generic
solar nebula models, but can easily exceed the disk column at larger
distances.  If the level of ionization is high enough--and the
Alfv\'en velocity $v_A$ is {\em small} enough---the gas within the range of
the cosmic rays will be MRI active.  (The Alfv\'en velocity cannot
be too large if disturbances are not to be stablized by magnetic tension.)
If these criteria are met, Gammie argued that turbulent accretion would proceed in
the outer layers of protostellar disks, but that the midplane regions
would remain laminar---in effect, a ``dead zone.''  This is the
basis of the concept of layered accretion, which has become an important
idea in protostellar disk modeling (Sano et al. 2000).

Gammie's original construction was based on the assumptions that the
accretion would occur in a layer of fixed column,
and that $\alpha$ is constant.  Taken together, these assumptions
preclude a steady solution; the mass flux rate is not independent
of position.  Instead, matter is deposited from the outer regions
into the disk's inner regions where it builds up.  At some point
in this scenario the disk become gravitationally unstable, and it
was speculated that an accretion outburst might occur, which
was tenatively identified
with FU Orionis behavior.  On the other hand, there is no compelling
argument (beyond mathematical convenience) to prevent the $\alpha$ parameter
from adjusting with position, if this allows a relaxation to
a time steady solution.  The overarching layered accretion picture
would, however, remain intact with this modification.  

YSO's are almost universally X-ray sources.  Glassgold et al. (2000)
noted that X-rays are potentially a far more powerful ionization source,
even if attenuated, than galactic cosmic rays.  These authors were thus
the first to draw the link between X-ray observations and MRI activity
of accretion disks.   In the Glassgold et al. formulation, X-rays
from a locally extended corona centered on the YSO irradiate the disk,
and depending upon the chosen parameters, could provide the requisite
ionization levels to lessen the extent of the dead zone or eliminate
it altogether.

Whether the dead zone persists in the presence of X-ray irradiation or not
depends, among other things, upon the model adopted for the disk structure.
Generally, the so-called minimum mass solar nebula model (Weidenschilling 1977)
is used.  This is a reconstruction of the surface density distribution
of the sun's protostellar disk based on current planetary masses and
compositions, and results in an $R^{-3/2}$ radial distribution.
Fromang, Terquem, \& Balbus (2002) suggested that ionization fractions
should be calculated self-consistently within the framework of accretion
disk theory, which in general leads to a more shallow dependence of 
surface density with radius.   By introducing  
accretion parameters whose values are free
($\alpha$ and the mass accretion rate $\dot M$), 
the range of parameter space increases, and the presence and extent
of a dead zone becomes yet more model dependent.  

The ionization fraction of a protostellar disk is a chemical problem, and
in principle can involve a very complex, uncertain, molecular reaction
network.  Ilgner \& Nelson (2004a) investigated the consequences for
the dead zone of a considerably richer chemical network.  While the
quantitative details were sensitive to the chemistry, the qualitative
structure was not.  A dead zone is likely to be present in any plausible
chemical network that has been studied up to the present, but there are
conditions in which its extent can be very small or possibly even zero.
Ilgner \& Nelson (2004b), for example, made the interesting comment that
flaring activity by the central star can have a significant effect on
the extent of the dead zone.

\subsection {Activity in the Dead Zone}

Just because the dead zone is unable to host the MRI does not mean that it
is well and truly dead.   Fleming and Stone (2003) carried out numerical
simulations in which the magnetically active upper layers of the disk
coupled dynamically to the magnetically inert dead zone, resulting in
a small but significant {\em Reynolds} stress, even though the Maxwell
stress was zero.  More recently, Turner \& Sano (2008) suggested that
the high resistivity characteristic of the dead zone is also a means
for magnetic field to diffuse into this region from the active layers.
Moreover, these authors point out, an active MRI region is not an absolute
prerequisite for accretion: direct magnetic torques on much larger scales
would also serve, and since they do not dissipate energy
in a turbulent cascade, are much less costly to maintain.

Terquem (2008) constructed global models of protostellar disks based on
the $\alpha$ prescription.  The presence of a dead zone was surprisingly
nondisruptive, provided that $\alpha$ was not too small.  Steady solutions
were found for values of $\alpha$ in the dead zone as small as $10^{-3}$
times the active zone value.  In these models, the dead zone was thicker
and more massive than its surroundings, but because of the relatively
weak $\alpha$ scalings, by less than an order of magnitude.
These results, taken as a whole, suggest that an embedded dead zone
in a protostellar disk is not necessarily disruptive to the accretion 
process.

\subsection {Dust}

In all of our discussions, we have been most negligent by
not mentioning the effects of small dust grains (e.g. Sano et al. 
2000).  Let us see crudely why this is
so.

Consider a mass $M$ of protostellar disk gas, of which $10^{-2}M$
is in the form of spherical dust grains of radius $r_d$.  If
the dust grains have a density of 3 gm cm$^{-3}$, there are a total
of
$$
N_d = (10^{-2}/4\pi)(M/r_d^3)
$$ 
grains in a volume $V$.
The grains present a geometric cross section of $\sigma_d=\pi r_d^2$
(we ignore here the enhancement due to the induced charge [Drain
\& Sutin 1987]), and the electrons have an average
radial velocity of some $v_e=1.6(kT/m_e)^{1/2}$.  The total
dust recombination rate per unit volume is
\beq
(N_d/V) n_e \sigma_d v_e
\eeq
where $n_e$ is the number density of electrons.  This should be 
compared with a typical dielectronic recombination rate of
$\beta\equiv 8.7\times 10^{-6}$ cm$^3$ s$^{-1}$(Glassgold, Lucas,
\& Omont 1986; Gammie 1996).  If $x$ is ionization fraction,
then the ratio of dielectronic recombination to dust recombination
is
\beq
{n_e \beta V\over N_d \sigma_d v_e} \sim \left(x\over 10^{-13}\right)
{r_d\over T}
\eeq
where $r_d$ is in cm and $T$ in K.  For small dust grains
($\sim 10^{-5}$cm) this ratio is $\ll 1$ and dust recombination
is overwhelmingly important.   But the relative importance of the grains
diminishes as they grow in size, especially in the cooler portions of the
disk.  Sano et al. (2000) concluded that in their model
(based on cosmic ray ionization) a typical protostellar disk
would be MRI stable inside of about 20 AU if small grain are present,
except for the innermost regions which would be thermally ionized.

The detailed effects of dust grains have been considered by many authors; a
very good review and list of references is given in the recent
paper of Salmeron \& Wardle (2008).  These authors have considered the
vertical structure of the magnetic coupling in a minimum mass solar
nebula model, including the effects of Hall electromotive fores and
ambipolar diffusion.  They employed a sophisticated chemical network
(Nishi et al. 1991), and small dust grains ranging in radius from 0.1 to
3 microns.  At the fiducial locations of 5 and 10 AU, Salmeron \& Wardle
found good magnetic coupling over an impressive range of magnetic field
strengths provided for the 3 micron grains, a considerable imporovement
from our naive estimate above.  For the 0.1 micron grains, the magnetic
coupling dropped sharply (as did MRI growth rates), and was restricted
to higher elevations above the disk plane.

\section {Summary}

Protostellar disks are gaseous systems dynamically dominated by Keplerian
rotation.  These disks are accreting onto the central protostar, so there
must be a source of enhanced angular momentum transport present.  In the
early stages of the disk's life, this enhanced transport may well be due
to self-gravity, with density waves largely responsible for moving angular
momentum outwards.  Once the disk becomes observable, its mass is below
the minimum needed to sustain self-gravitational spiral wave transport.
The only established mechanism able to sustain high levels of angular
momentum transport is MHD turbulence produced by the magnetorotational
instability, or MRI.

The dynamical effects of magnetic fields on protostellar disks depends
crucially on the degree of ionization, more specifically on the electron
fraction, that is present.  This means that there is a direct link
between the gross dynamical behavior of a protostellar disk and its
detailed chemical profile.   In principle, the full multifluid nature
of the disk gas --- neutrals, ions, electrons, and dust grains ---
must be grappled with at some level to elicit and understand the disk
strucutre.  Fortunately, only a very small ionization level is needed
to couple the charged and neutral components, leaving (in essence),
a single magnetized fluid.   An electron fraction of $10^{-13}$ is a
typical fiducial number for the threshold of magnetic coupling at 1 AU
in the T Tauri phase of the solar nebula: roughly one electron per cubic
millimeter of disk gas!  Unfortunately, however, on scales of 10's of
AU, near the dense midplane the ionization of protostellar disks may
not even rise to this minimal level of ionization.  At the boundary
between good and poor magnetic coupling, the MHD processes are complex,
not well-understood, and the domain of ongoing inquiry.  

In regions of the disk where the magnetic coupling is sufficient, the
combination of a weak (subthermal) magnetic field and Keplerian rotation
leads to the MRI.  In ideal MHD, the magnetic field behaves as though
it were frozen into the conducting gas, and the presence of differential
rotation produces azimuthal magnetic field from a radial magnetic field.
But this is not all that happens.  If one tries to simulate this simple
process on a computer, the laminar flow breaks down into a turbulent mess,
even if the field is very weak.

The reason for this is due to another classical property of magnetic
fields, that the force exerted by the lines of force on the background
gaseous fluid is the sum of a pressure-like term and a tension-like term.
It is the latter tension-like term that is important for an understanding
of the MRI.  When two nearby fluid elements are moved apart, even if
only because of a random perturbation, the magnetic tension force acts
precisely like a spring coupling the two masses.  The fluid element on
the inside rotates faster than the element on the outside, and tries to
speed it up.  The outer element, in acquiring angular momentum from the
inner element, finds itself too well-endowed, and spirals outward toward
a higher orbit where its excess angular momentum can be accommodated.
On the other hand, the inner element, having lost angular momentum, finds
itself at a deficit and must drop to a lower angular momentum orbit.
This separation stretches the field lines that couple the elements,
the magnetic tension goes up, and the process runs away.   This is
the MRI.  Notice that angular momentum transport is at the very core of
the linear instability, rather than the result of some sort of nonlinear
mixing process.

But the MRI does, in fact, lead to rapid turbulent mixing of the disk
gas, as outwardly moving and inwardly moving fluid elements encounter
one another and dissipate their energy.  If this happens locally, then
the ingredients for a classical $\alpha$ model of accretion are
present.  To the extent that many disk features are insensitive to,
or independent of, the precise rms level of the disk fluctuations,
these models can be of some practical utility.  The classical formula
relating the disk emissivity to radius $R$ is the most important
instance of this.  

In real protostellar disks, the level of ionization present is such
that there are significant departures from the behavior of an ideal
(perfectly conducting) MHD gas.  In principle, the gas can become
completely decoupled from the magnetic field, and go over to a
hydrodynamic system.  Less dramatically, the current-bearing electrons
can acquire a velocity significantly different from the dominant neutrals,
since the former need to maintain a current density and magnetic field
even as charged species become rarer and rarer.  When this happens, 
the field lines are no longer frozen into the motion of the bulk
of the (neutral) disk gas, they are frozen into the electrons
and the distinction becomes important.    In the so-called Hall regime,
the ions and neutrals move together distinctly from the electrons,
and the ion motion relative to the electron-following field lines
induces an additional electromotive force into the gas, beyond
the self-induction responsible for simple field line freezing.  
Hall MHD is likely to be important in protostellar disks on scales
of AU to 10's of AU scales.  

Dust grains are an ever present complication for MHD disk modeling.
Typically, solids make up about 1 percent of the mass of interstellar
gas.  An interstellar population of small grains (radius $\sim 10^{-5}$ cm)
with a total mass fraction of a percent would present an enormous
collecting area of would-be gas phase electrons.  Putting charges on 
the grains effectively removes them because of the low mobility of
the grains.  Determining the disposition of the dust grains is thus
a necessity for constructing MHD disk models.  As the disk evolves, so
too do the dust grains.  They grow in size as they agglomerate, and they
tend to settle towards the midplane, if they are not stirred by turbulence
or some other dynamical process.  Larger dust grains are much less 
efficient in removing gas phase electrons than are smaller grains,
where ``larger'' means growth in radius of an order of magnitude
or more.  

Where does all of this leave us?   Clearly, we are a long way from
framing a picture of protostellar disk evolution at the level of, say,
classical stellar evolution.  But for all of the gaping uncertainties,
a useful zeroth order MHD-based picture of a protostellar disk in its
T Tauri phase can be cobbled together:

Inside of about $0.3$ AU, a protostellar disk will be thermally ionized
by direct exposure to the central source and self-consistently by the
dissipation of MHD turbulence.  We may expect vigorous MRI induced MHD
turbulence in this zone.

On scales of AU's, thermal ionization is no longer adequate to maintain
the requiste levels of ionization to ensure MHD coupling, at least not
near the high density midplane.  Depending upon the X-ray luminosity
of the central star, the spectrum of dust grains and the abundance of
gas phase potassium and sodium (electron donors), there could be good
coupling at low density, higher elevation disk altitudes.  The ``dead
zone'' at lower altitudes need not be devoid of all transport; density
waves from an adjacent active layer or large scale magnetic torques
would each contribute their own stresses.  It is even possible that
some degree of coupling could be maintained down to the disk midplane,
though this depends strongly on modeling assumptions.   It seems likely,
however, that on a scale an AU to tens of AU's , the level of MHD turbulence
will be far less than in the disk's innermost regions.

Finally, on scales larger than tens of AU's, the falling density
lengthens recombination times and suggests a return to good MHD coupling.
The typical size of a disk is many hundred AU's so that the dead zone is
small when viewed globally, and may not have much of an impact in the
overall accretion process (Terquem 2008).  Perhaps, however, it is not
coincidence that the MHD ``quiet zone'' coincides with the region of
planet formation in the solar nebula and (more speculatively) in other
protostellar disks.

\section*{Acknowledgements}

I would like to thank my collaborators over past several years with
whom and from whom I have learned much of what I know of the subject of
protostellar disks: O. Blaes, S. Fromang, C. Gammie, J. Hawley, J. Stone,
and C. Terquem.  I would also like to thank specifically P. Garcia and
the referee for their important respective comments on the H-H$^+$
reaction rates and ionization equilibrium.  This work was supported
by a Chaire d'Excellence from the French Ministry of Higher Education,
and a grant from the Conseil R\'egional de l'Ile de France.


\section*{Appendix A}

Begin with equation (\ref{ambi}):
\beq
{ \bb{J}\over c}\btimes \bb{B} = \bb{p_{In}} + \bb{p_{en}}
\eeq
where the right side of this equation is
\beq\label{appp}
n_I n \mu_{In}\langle \sigma_{nI}w_{nI}\rangle (\vi-\vv)+
n_e n m_{e}\langle \sigma_{ne}w_{ne}\rangle [(\ve-\vi) + (\vi-\vv)]
\eeq
In what follows, we will often need an estimate of the ratio
of the ion and electron collision rates.  Following our discussion 
in section 3, we will assume that
\beq
{\langle \sigma_{nI}w_{nI}\rangle \over
\langle \sigma_{ne}w_{ne}\rangle } = \left(  {m_e\over \epsilon\mu_{In}}
\right)^{1/2}
\eeq
where $\epsilon < 1$ is inserted because the electron-neutral
cross section is geometrical, while the ion-neutral collision cross section
is larger than geometrical.  

In (\ref{appp}),
the first $(\vi-\vv)$ term dominates the last by a factor of order
$(\mu_{In}/\epsilon m_{e})^{1/2}$.  Hence,
\beq
\bb{p_{In}} = 
{ \bb{J}\over c}\btimes \bb{B} -\bb{p_{en}} =      
{ \bb{J}\over c}\btimes \bb{B}  + {nm_e\over e}
\langle \sigma_{ne}w_{ne}\rangle \bb{J}
\eeq
With the help of equation (\ref{pinn}), this implies
\beq\label{ap1}
\vi-\vv = {\bb{J\times B}\over c\gamma\rho\rho_I} +
\underbrace{\sqrt{\epsilon m_e\over\mu_{nI}}Z
(\vi-\ve)}_A.
\eeq
We have marked the term with an ``A'' for future reference.

Recall equation (\ref{vees}):
\beq\label{app2}
\bb{E}+
{1\over c} \left[
\vv +(\ve-\vi)+
(\vi-\vv) \right] \btimes\bb{B} + {m_e \nu_{en}\over e}
\left[ (\ve-\vi) +(\vi -\vv) \right]=0.
\eeq
In substituting equation(\ref{ap1}) for $\vi-\vv$ in the
above, we may always drop the ``A'' term, since it is small
compared with $\ve-\vi$.

Proceeding with the above substitution leads to
\beq
\bb{E}+
{\vv\over c}\btimes{B}- {\bb{J\times B}\over en_e c}
\bigg[ 1 -\underbrace{{m_e\nu_{en} n_e\over \gamma\rho\rho_I}}_B\bigg]
+{(\bb{J\times B})\btimes\bb{B}\over c^2\gamma\rho\rho_I} -{\bb{J}\over
\sigma_{cond}} =0,
\eeq
where $\sigma_{cond}$ is defined in equation (\ref{cond1}).  The ``B'' term
in the above equation
may now clearly be dropped: it is of order $Z(\epsilon m_e\mu_{nI})^{1/2}/m_n$.
This leads to
\beq
\bb{E}+
{\vv\over c}\btimes{B}- {\bb{J\times B}\over en_e c}
+{(\bb{J\times B})\btimes\bb{B}\over c^2\gamma\rho\rho_I} -{\bb{J}\over
\sigma_{cond}} =0,
\eeq
which is precisely equation (\ref{ohm1}) in the text.

\section* {Appendix B:}

To estimate the order of departures from charge neutrality or
the displacement currents, we will assume that the $\del$ operator
is $\sim 1/L$, where $L$ is a characteristic length scale of the flow,
and $\dd/\dd t$ is $\sim v/L$, where $v$ is a characteristic
velocity (say, the largest of the neutral, ion, or electron velocities).

For the electric field, we take $E\sim vB/c$, since we are only
interested in problems where the inductive terms are comparable to,
or in larger than, the resistive damping.
Then
\beq
\del\bcdot\EE \sim E/L\sim vB/Lc\sim 4\pi vJ/c^2 \sim 4\pi e n_e v^2/c^2
\eeq
Hence, the divergence of $\EE$ is of order $v^2/c^2$ times the electron
charge density (at most, since we assumed $v_I-v_e \sim v$ in the above).
It may thus be ignored.

For the displacement current, the demonstration is almost a matter of
direct inspection.  If we take the curl of equation (\ref{displace})
and use equation (\ref{faraday}), then the on the right
side of (\ref{displace}) the first term is $\sim B/L^2$, while
the second, displacement, term is $\sim v^2  B/c^2$.
It may thus be ignored.


\end{document}